\documentclass[article]{jss}

\usepackage{thumbpdf,lmodern}

\usepackage{amsmath, amsfonts,amssymb}
\usepackage{mathtools}
\usepackage{framed}
\usepackage{changebar}
\usepackage{graphicx}
\usepackage{comment}

\DeclarePairedDelimiter\abs{\lvert}{\rvert}%
\newcommand{\var}{\operatorname{var}}

\newcommand{\diag}{\operatorname{diag}}
\newcommand{\vect}{\operatorname{vec}}
\newcommand{\plim}{\operatorname{plim}}

\newcommand{\myincfig}[3]{%
	\begin{figure}[h]
	\centering
	\resizebox{0.85 #2}{!}{\includegraphics{#1}}
	\caption{\label{#1} #3}
	\end{figure}
	}

\newcommand{\bcb}{\begin{changebar}}
\newcommand{\ecb}{\end{changebar}}

\newtheorem{theorem}{Theorem}
\newtheorem{proposition}{Proposition}

\author{Marina Knight\\University of York
   \And Kathryn Leeming\\University of Warwick\\
   \And Guy Nason\\Imperial College London\\
   \And Matthew Nunes\\University of Bath}
\Plainauthor{Marina Knight, Kathryn Leeming, Guy Nason, Matthew Nunes}

\title{Generalised Network Autoregressive Processes and the \pkg{GNAR} package}
\Plaintitle{Generalised Network Autoregressive Processes and the GNAR package}
\Shorttitle{GNAR processes}

\Abstract{
This article introduces the \pkg{GNAR} package, which fits, predicts, and simulates from a powerful new class of generalised network
autoregressive processes. Such processes consist of a multivariate
time series along with a real, or inferred, network that provides information
about inter-variable relationships. The GNAR model relates values of a time series
for a given variable and time to earlier values of the same variable and of neighbouring variables, with inclusion
controlled by the network structure.
The \pkg{GNAR} package is designed to fit this new model, while working with standard \code{ts} objects and the \pkg{igraph} package for ease of use.
}

\Plainkeywords{multivariate time series, networks, missing data, network time series}

\Address{
  Marina Knight\\
  Department of Mathematics\\
  University of York, UK\\
   E-mail: \email{marina.knight@york.ac.uk}\\
  URL: \url{https://www.york.ac.uk/maths/staff/marina-knight/}

    \emph{and}\\

  Kathryn Leeming\\
  Department of Statistics\\
  University of Warwick, UK\\
  E-mail: \email{kathryn.leeming@warwick.ac.uk}\\
  URL: \url{https://www.warwick.ac.uk/fac/sci/statistics/staff/academic-research/leeming/}

  \emph{and}\\

  Guy P.\ Nason\\
  Department of Mathematics\\
  Imperial College London, UK\\
  E-mail: \email{g.nason@imperial.ac.uk}\\
  URL: \url{http://www.imperial.ac.uk/people/g.nason}

  \emph{and}\\

  Matthew Nunes\\
  School of Mathematical Sciences\\
  University of Bath, UK.
  E-mail: \email{m.a.nunes@bath.ac.uk}\\
  URL: \url{http://people.bath.ac.uk/man54/}\\

}

\begin{document}



\section{Introduction}
Increasingly within the sciences, networks and network methodologies are being used to answer research questions. Such networks might be observed, such as connections
 in communication network or information flows within, or they could be unobserved:
 inferred networks that can explain a process or effect.
 Given the increase in the size of data sets,
it may also be useful to infer a network from data to efficiently summarise the data generating process.

We consider time series observations recorded at different nodes of a network, or graph. Our
\pkg{GNAR} package \citep{leeming18:GNAR} and its novel generalised network autoregressive (GNAR) statistical models
focus on partnering a network with a multivariate
time series and modelling them jointly. One can find an association network, see,
e.g., Chapter~7 of \cite{Kolaczyk2009},
or Granger causality network, e.g., \cite{Dahlhaus2003},
between different variables by analysing a multivariate time series and its properties.
However,
here we assume the existence of an underlying network and use it during the analysis of the time series, although sometimes its complete structure is unknown.

Networks can provide
strong information about the dependencies between variables.
Within our generalised network autoregressive (GNAR) model, each node depends on its previous values as in the univariate autoregressive
framework, but also may depend on the previous values at its neighbours, neighbours of neighbours, and so on. Our GNAR modelling framework is flexible, allowing for different types of network,
networks that change their structure over time (time-varying networks),
 and also can be powerfully applied in the important practical
 situation where the time series feature missing
 observations.

Driven in part by the increased popularity and recent research activity in the field of statistical network analysis, there has been a concurrent growth in software for analysing such data.  An exhaustive list of these packages is beyond the scope of this article, but we review some relevant ones here.

Existing software in this area predominantly focusses on the various models for network-structured data.  In the static network setting, these include packages dedicated to latent space network models, such as \pkg{collpcm} \citep{wyse17:collpcm}, \pkg{HLSM} \citep{adhikari18:HLSM}, \pkg{latentnet} \citep{krvitsky18:latentnet} amongst others; exponential random graph models and their variants, for example \pkg{ergm} \citep{handcock18:ergm}, \pkg{GERGM} \citep{denny18:GERGM} or \pkg{hergm} \citep{schweinberger18:hergm}; and block models in e.g., \pkg{blockmodels} \citep{leger15:blockmodels}.  For dynamic networks, packages for time-varying equivalents of these network models are also available, see e.g., the \pkg{tergm} package \citep{krvitsky18:tergm} or \pkg{dynsbm} \citep{matias18:dynsbm}.  There are also a multitude of more general packages for network analysis, e.g., for network summary computation or implementations of methodology in specific applications of interest.

Despite this, software dedicated to the analysis of time series and other \emph{processes} on networks is sparse.  A number of packages implement epidemic (e.g., SIR) models of disease spread, notably \pkg{epinet} \citep{groendyke18:epinet}, \pkg{EpiLM}/\pkg{EpiLMCT} \citep{warriyar18:EpiLM, almutiry18:EpiLMCT} and \pkg{hybridModels} \citep{marquez18:hybridModels}; these use transmission rates to model processes as opposed to temporal and network dependence through time series models as in \pkg{GNAR}.  Similarly, the \pkg{NetOrigin} software \citep{manitz18:NetOrigin} is dedicated to source estimation for propagation processes on networks, rather than fitting time series models.  Packages such as \pkg{networkTomography} \citep{blocker14:networkTomography} deal with time-varying models for (discrete) count processes or flows on links of a \emph{fixed} routing network;  the \pkg{tnam} package \citep{leifeld17:tnam} fits models using \emph{spatial} (and not network-node) dependence.  Both of these are in contrast to the \pkg{GNAR} package,
which implements time series models which account for known \emph{time-varying network structures}.

Other packages can implicitly develop network-like structured time series models through penalised or constrained variable selection, such as \pkg{autovarCore}
\citep{emerencia18:autovarCore}, \pkg{nets} \citep{brownlees17:nets}, \pkg{sparsevar} \citep{vazoller16:sparsevar}, as well as the \pkg{vars} package \citep{Pfaff2008}.
Packages that take a graphical modelling approach to the dependence structure within time series include \pkg{gimme} \citep{lane19:gimme}, \pkg{graphicalVAR}
\citep{epskamp18:graphicalVAR}, \pkg{mgm} \citep{haslbeck19:mgm}, \pkg{mlVAR} \citep{epskamp19:mlVAR}, and \pkg{sparseTSCGM} \citep{abegaz16:sparseTSCGM}.  These approaches also differ fundamentally from the GNAR models since the network is constructed during analysis, as opposed to \pkg{GNAR},
which specifically incorporates information on the network structure into the model \emph{a priori}.  The \pkg{vars} package features in Section~\ref{sec:results}, where we highlight the differences between the GNAR models and this existing class of techniques.

Section~\ref{Model} introduces our model, and demonstrates how \pkg{GNAR}
can be used to fit network models to simulated
network time series in Section~\ref{exsim}.
Order selection and prediction are discussed in Section~\ref{Implementation}, which includes
an example of how to use BIC to select model order for a wind speed network
time series in Section~\ref{wind}.
An extended example, concerning constructing a network to aid GDP forecasting,
is presented in Section~\ref{gdp}.
Section~\ref{Discussion} discusses different network modelling options that could be chosen, and presents a summary of the article.
All results were calculated using version~3.5.1 of the statistical software \proglang{R} (\cite{Rcore2017}).
\section{Network time series processes} \label{Model}
We assume that our multivariate time series follows an autoregressive-like model at each node, depending both on the previous values of the process at that node, and on neighbouring nodes at previous time steps. These neighbouring nodes are included as part of the network structure, as defined below.
\subsection{Network terminology and notation}
Throughout  we assume the presence of one or more networks, or graphs, associated with the observed time series. Each univariate time series that makes up the multivariate time series occurs, or is observed at,
a node, or location on the graph(s). These nodes are connected by a set of edges, which may be directed, and/or weighted.

We denote a graph by $\mathcal{G} = (\mathcal{K},\mathcal{E})$, where $\mathcal{K}=\{1,...,N\}$ is the set of nodes, and $\mathcal{E}$ is the set of edges. A directed edge from node $i \in \mathcal{K}$ to $j \in \mathcal{K}$ is denoted $i \rightsquigarrow j$, and an un-directed edge between the nodes is denoted $i \leftrightsquigarrow j$. The edge set of a directed graph is $\mathcal{E} = \{ (i,j): i \rightsquigarrow j; i,j \in \mathcal{K} \}$, and similarly for the set of un-directed edges.
\subsubsection{Stage-$r$ neighbourhoods}
We introduce the notion of neighbours and stage-neighbours in the graph structure as follows; for a subset $A \subset \mathcal{K}$ the neighbour set of $A$ is given by $\mathcal{N}(A) = \{j \in \mathcal{K}/A: i \rightsquigarrow j ; i \in A \}$. These are the first neighbours, or stage-1 neighbours of $A$. The $r$th stage neighbours of a node $i \in \mathcal{K}$ are given by $\mathcal{N}^{(r)}(i) = \mathcal{N} \{ \mathcal{N}^{(r-1)}(i) \} / [ \{\cup_{q=1}^{r-1} \mathcal{N}^{(q)}(i)\} \cup \{i\}] $, for $r=2,3,...$ and $\mathcal{N}^{(1)}(i) = \mathcal{N}(\{i\})$.

Figure~\ref{jss3594-pltnet} shows an example graph, where node E has stage-1 neighbour A, stage-2 neighbour D, and stage-3 neighbours B and C. Neighbour sets for this example include $\mathcal{N}^{(1)}(D)= \{A, B, C\}$, and $\mathcal{N}^{(3)}(E)= \{B, C \}$. In the time-varying network setting, a subscript $t$ is added to the neighbour set notation.


\subsubsection{Connection weights} \label{connectionweights}
Each network can have connection weights $\omega \in [0,1]$ associated with every pair of nodes.
This connection weight can depend on the size of the neighbour set and also encodes any edge-weight information.
Formally, the values of the connection weights from a node $i \in \mathcal{K}$ to its stage-$r$ neighbour $j \in \mathcal{N}^{(r)}(i)$ will be the reciprocal of the number of stage-$r$ neighbours; $\omega_{i,j} = |\mathcal{N}^{(r)}(i) |^{-1} $, where $|\cdot |$ denotes the cardinality of a set. In Figure~\ref{jss3594-pltnet} the connection weights would be, for example, $\omega_{E,A} = 1$, $\omega_{A, E}=\omega_{A,D}=0.5$.
Connection weights are not necessarily symmetric, even for an un-directed graph.
We note that this choice of these inverse distance weights is one of many possibilities,
and some other
means of creating connection weights could be used.

When the edges are weighted, or have a distance associated with them, we use the concept of distance to find the shortest path between two vertices. Let the distance from node $i$ to $\ell$ be denoted $d_{i,\ell} \in \mathbb{R}_{+}$, and if there is an un-normalised weight between these nodes, denote this $\mu_{i,\ell}\in \mathbb{R}_{+}$. To find the length of connection between a node $i$ and its stage-$r$ neighbour, $k$, we sum the distances on the paths with $r$ edges from $i$ to $k$ and take the minimum (note that there are no paths with fewer edges than $r$ as $k$ is a stage-$r$ neighbour). If the network includes weights rather than distances, we find the shortest $r$ length path where $d_{i,\ell}=\mu_{i,\ell}^{-1}$. Then the connection weights between node $i$ and its stage-$r$ neighbour $k$ are either $\omega_{i,k} = d_{i,k}^{-1} \{ \sum_{\ell \in \mathcal{N}^{(r)}(i)} d_{i,\ell}^{-1} \}^{-1}$ for distances, or $\omega_{i,k} = \mu_{i,k} \{ \sum_{\ell \in \mathcal{N}^{(r)}(i)} \mu_{i,\ell} \}^{-1}$ for a network with weights. This definition means that all nodes will have connection weights that sum to one for any non-empty neighbour set, whether they are in a sparse or dense part of the graph.
\subsubsection{Edge or node covariates}
A further  important innovation permits
a covariate that can be used to encode edges effects (or nodes) into certain types.
Our covariate will take $C \in \mathbb{N}$ discrete values and be indexed by $c$. A more general covariate could be
considered, but we wish to keep our notation simple in the definition that follows. For example,
in an epidemiological network we might have two edge types:
one that carries information about windborne spread of
infection and the other carries information about identified direct infections.
The covariates do not change our neighbour sets or connection weight definitions,
so we have the property $\sum\limits_{q \in \mathcal{N}^{(r)}(i)} \sum\limits_{c=1}^C \omega_{i,q,c} = 1$ for all $i \in \mathcal{K}$ and $r \in \mathbb{N}$ such that $\mathcal{N}^{(r)}(i)$ is non-empty.

\subsection{The generalised network autoregressive model}\label{NARmod}
Consider an $N\times 1$ vector of nodal time series, $\mathbf{X}_t=(X_{1,t},\hdots,X_{N,t})'$, where $N$ is considered fixed.  Our aim is to model the dependence structure within and between the nodal series using the network structure provided by (potentially time-varying) connection weights, $\omega$.
For each node $i \in \{1,\hdots,N\}$ and time $t\in \{1,\hdots,T\}$, our generalised autoregressive model of order $(p,[\mathbf{s}])\in \mathbb{N}\times\mathbb{N}_0^p$
for $\mathbf{X}_t$ is
\begin{equation} \label{NAR1}
    X_{i,t} = \sum_{j=1}^p \left( \alpha_{i,j} X_{i,t-j} +\sum_{c=1}^C \sum_{r=1}^{s_j} \beta_{j,r, c}\sum_{q \in \mathcal{N}^{(r)}_t(i)} \omega_{i,q,c}^{(t)} X_{q,t-j} \right) + u_{i,t},
\end{equation}
where $p\in\mathbb{N}$ is the maximum time lag, $[\mathbf{s}]=(s_1,\hdots,s_p)$ and $s_j \in \mathbb{N}_0$ is the maximum stage of neighbour dependence for time lag $j$, with $\mathbb{N}_0 = \mathbb{N}\cup \{0\}$,
 $\mathcal{N}^{(r)}_t(i)$ is the $r$th stage neighbour set of node $i$ at time $t$, $\omega_{i,q,c}^{(t)} \in [0,1]$ is the connection weight between node $i$ and node $q$ at time $t$ if the path corresponds
to covariate $c$. Here, we consider a sum from one to zero to be zero, i.e., $\sum_{r=1}^0 ( \cdot ) \coloneqq 0$.
The $\alpha_{i, j} \in \mathbb{R}$ are `standard' autoregressive parameters at lag $j$ for node $i$.
The $\beta_{j, r, c} \in \mathbb{R}$ correspond to the effect of the $r$th stage neighbours, at lag $j$, according to covariate
$c= 1, \ldots, C$. Later, we derive conditions on the model parameters to achieve process stationarity over the network.
Here the noise, $\{u_{i,t}\}$, is assumed to be independent and identically distributed at each node $i$, with mean zero and variance $\sigma_i^2$.
Our model  meaningfully enhances that of the arXiv publication
\cite{Knight2016} by now additionally including
different autoregressive parameters,
connection weights at each node
and, particularly, parameters $\beta$ that depend on covariates.  Note that the IID assumption on the noise $\{u_{i,t}\}$ could of course be relaxed to include correlated innovations. 

We note that crucially, the time-dependent network topology is integral to the model parametrisation through the use of time-varying weights and neighbours. These features yield a model that is sensitive to the network structures and captures contemporaneous as well as autoregressive relationships, as defined by equation~\eqref{NAR1}. The network should therefore be viewed not as an estimable quantity, but as a time-dependent known structure.

In the  GNAR model, the network may change over time, but the covariates stay fixed. This means that the underlying network can be altered over time,
for example, to allow for nodes to drop in and out of the series but model fitting can still be carried out. Practically, this is extremely useful, as shown by the example in Section~\ref{gdp}.
Our model allows for the $\alpha$ parameters may be different at each node,
however the interpretation of the network regression parameters, $\beta_{j, r, c}$, is the same throughout the network.

A more restrictive version of the above model is the global-$\alpha$ GNAR$(p, [\textbf{s}])$ model, which has the same autoregressive covariate at each node,
where the $\alpha_{i,j}$ are replaced by $\alpha_j$.
This defines a process with the same behaviour at every node, with differences being present only due to the graph structure.

\subsection{GNAR network example}

Networks in the \pkg{GNAR} package are stored in a list with two components
\code{edges} and \code{dist}. The \code{edges} component is itself a list with $N$ slots
each containing a vector whose entries are indices to their neighbouring nodes. For example,
if $3 \leftrightsquigarrow 4$ denotes an undirected edge between nodes $3$ and $4$ then
the vector \code{edges[[3]]} will contain a \code{4} and \code{edges[[4]]} will contain a \code{3}.
If the network is undirected this will mean that each edge is `double counted' in summary information. A directed edge $3 \rightsquigarrow 4$ would be listed in
\code{edges[[3]]} as a \code{4}, but not \code{edges[[4]]} if there is no edge in the opposite direction.
The \code{dist} component is of the same format as \code{edges}, and contains the distances corresponding to the edge links, if they exist.
For example, in an un-weighted setting, the connection weights are such that all neighbours of a node have equal effect on the node. This is achieved by setting all entries of the \code{dist} component to one, and the software calculates the connection weight from these.
A \pkg{GNAR}  network is stored in a \code{GNARnet} object, and an object
can be checked using the \code{is.GNARnet} function.
The S3 methods \code{plot}, \code{print}, and \code{summary} are available for \code{GNARnet} objects.

Figure~\ref{jss3594-pltnet} shows an example that is stored as a \code{GNARnet}
object called \code{fiveNet} and can be reproduced using
\begin{Schunk}
\begin{Sinput}
R> library("GNAR")
R> library("igraph")
R> plot(fiveNet, vertex.label = c("A", "B", "C", "D", "E"))
\end{Sinput}
\end{Schunk}
\myincfig{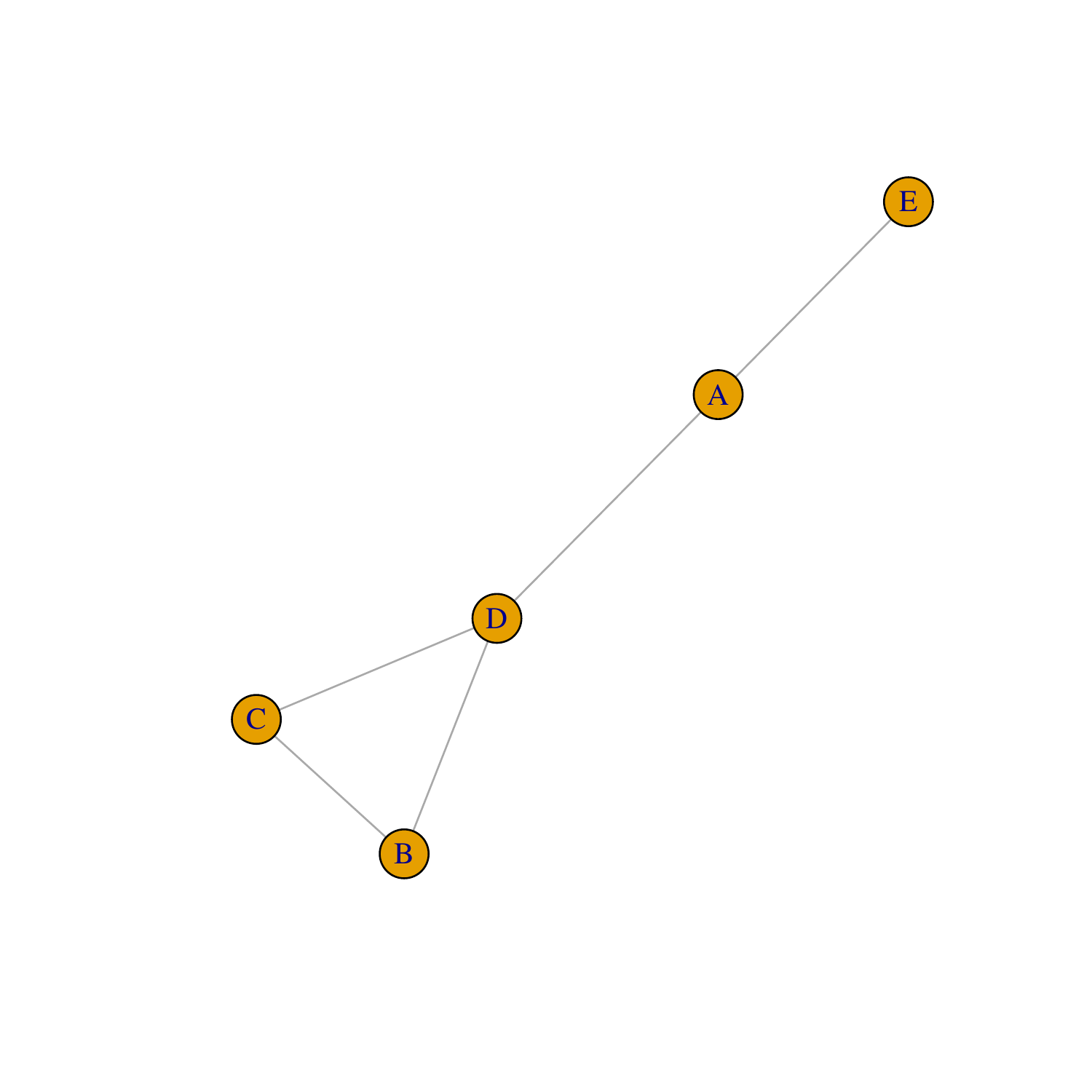}{\textwidth}{An example un-directed, un-weighted graph with five nodes labelled A to E.}

The basic structure of the \code{GNARnet} object is, as usual, displayed with
\begin{Schunk}
\begin{Sinput}
R> summary(fiveNet)
\end{Sinput}
\begin{Soutput}
GNARnet with 5 nodes and 10 edges
 of equal length  1
\end{Soutput}
\end{Schunk}
\subsubsection{Converting a network to GNARnet form} \label{GNARnet}
Our \code{GNARnet} format integrates with other methods of specifying a network via a set of functions that generate a \code{GNARnet} from others, such as an
\code{igraph} object.

An \code{igraph} object can be converted to and from the \code{GNARnet} structure using the functions \code{igraphtoGNAR} and \code{GNARtoigraph}, respectively.
For example, starting with the \code{fiveNet} \code{GNARnet} object,
\begin{Schunk}
\begin{Sinput}
R> fiveNet2 <- GNARtoigraph(net = fiveNet)
R> summary(fiveNet2)
\end{Sinput}
\begin{Soutput}
IGRAPH 41ddef8 U-W- 5 5 -- 
+ attr: weight (e/n)
\end{Soutput}
\begin{Sinput}
R> fiveNet3 <- igraphtoGNAR(fiveNet2)
R> all.equal(fiveNet, fiveNet3)
\end{Sinput}
\begin{Soutput}
[1] TRUE
\end{Soutput}
\end{Schunk}
whereas the reverse conversion would be performed as
\begin{Schunk}
\begin{Sinput}
R> g <- make_ring(10)
R> print(igraphtoGNAR(g))
\end{Sinput}
\begin{Soutput}
GNARnet with 10 nodes 
edges:1--2 1--10 2--1 2--3 3--2 3--4 4--3 4--5 5--4 5--6 
     6--5 6--7 7--6 7--8 8--7 8--9 9--8 9--10 10--1 10--9 
     
 edges of each of length  1 
\end{Soutput}
\end{Schunk}
We can also use the \code{GNARtoigraph} function to extract graphs
involving higher-order neighbour structures,
for example, creating a network of third-order neighbours.

In addition to interfacing with \code{igraph}, we can convert between \code{GNARnet} objects and adjacency matrices using functions \code{as.matrix} and
\code{matrixtoGNAR}.  We can produce an adjacency matrix for the \code{fiveNet} object with
\begin{Schunk}
\begin{Sinput}
R> as.matrix(fiveNet)
\end{Sinput}
\begin{Soutput}
     [,1] [,2] [,3] [,4] [,5]
[1,]    0    0    0    1    1
[2,]    0    0    1    1    0
[3,]    0    1    0    1    0
[4,]    1    1    1    0    0
[5,]    1    0    0    0    0
\end{Soutput}
\end{Schunk}
and an example converting a weighted adjacency matrix to a \code{GNARnet} object is
\begin{Schunk}
\begin{Sinput}
R> adj <- matrix(runif(9), ncol = 3, nrow = 3)
R> adj[adj < 0.3] <- 0
R> print(matrixtoGNAR(adj))
\end{Sinput}
\begin{Soutput}
GNARnet with 3 nodes 
edges:1--1 1--3 2--2 3--1 3--2 
 edges of unequal lengths 
\end{Soutput}
\end{Schunk}

\subsection{Example: GNAR model fitting}\label{exsim}
The \code{fiveNet} network has a simulated multivariate time series associated
with it of class \code{ts} called \code{fiveVTS}.
The pair together are a network time series.  The object can be loaded in the usual way using the \code{data} function.
 \pkg{GNAR} contains functions for fitting and predicting from GNAR models: \code{GNARfit} and the \code{predict} method, respectively. These make use of the familiar \proglang{R} command \code{lm}, since the GNAR model can be essentially re-formulated as a linear model, as we shall see in Section~\ref{Implementation} and Appendix~\ref{consistency}.  As such, least squares variance / standard error computations are also readily obtained, although other, e.g., HAC-type variance estimators could also be considered for GNAR models.

Suppose we wish to fit the global-$\alpha$ network
time series model GNAR$(2, [1, 1])$, a model with four parameters in total.
We can fit this model with the following code.
\begin{Schunk}
\begin{Sinput}
R> data("fiveNode")
R> answer <- GNARfit(vts = fiveVTS, net = fiveNet, alphaOrder = 2,
+    betaOrder = c(1, 1))
R> answer
\end{Sinput}
\begin{Soutput}
Model: 
GNAR(2,[1,1]) 

Call:
lm(formula = yvec ~ dmat + 0)

Coefficients:
 dmatalpha1  dmatbeta1.1   dmatalpha2  dmatbeta2.1  
    0.20624      0.50277      0.02124     -0.09523  
\end{Soutput}
\end{Schunk}
In this fit, the global autoregressive parameters are  $\hat{\alpha}_1 \approx 0.206$ and
$\hat{\alpha}_2 \approx 0.021$ and the $\beta$ network parameters are
$\hat{\beta}_{1,1, 1} \approx 0.503$ and $\hat{\beta}_{2,1, 1} \approx -0.095$. Also,
the network edges only have one type of covariate so $C = c=1$.
We can just look at one node. For example, the model at node~A is
\begin{displaymath}
X_{A,t} = 0.206X_{A,t-1} + 0.503 (X_{E,t-1} + X_{D, t-1})/2
  + 0.021X_{A, t-2} - 0.095 (X_{E,t-2} + X_{D, t-2})/2 + u_{E,t}.
\end{displaymath}
The model coefficients can be extracted from a \code{GNARfit} object
 using the \code{coef} function as is customary.
The \code{GNARfit} object returned by \code{GNARfit} function also has methods to extract fitted values and the residuals.
For example, Figure~\ref{jss3594-gnarfvp} shows the  first node time series and the
residuals from fitting the model. Figure~\ref{jss3594-gnarfvp} was produced by
\begin{Schunk}
\begin{Sinput}
R> plot(fiveVTS[, 1], ylab = "Node A Time Series")
R> lines(fitted(answer)[, 1], col = 2)
\end{Sinput}
\end{Schunk}
\myincfig{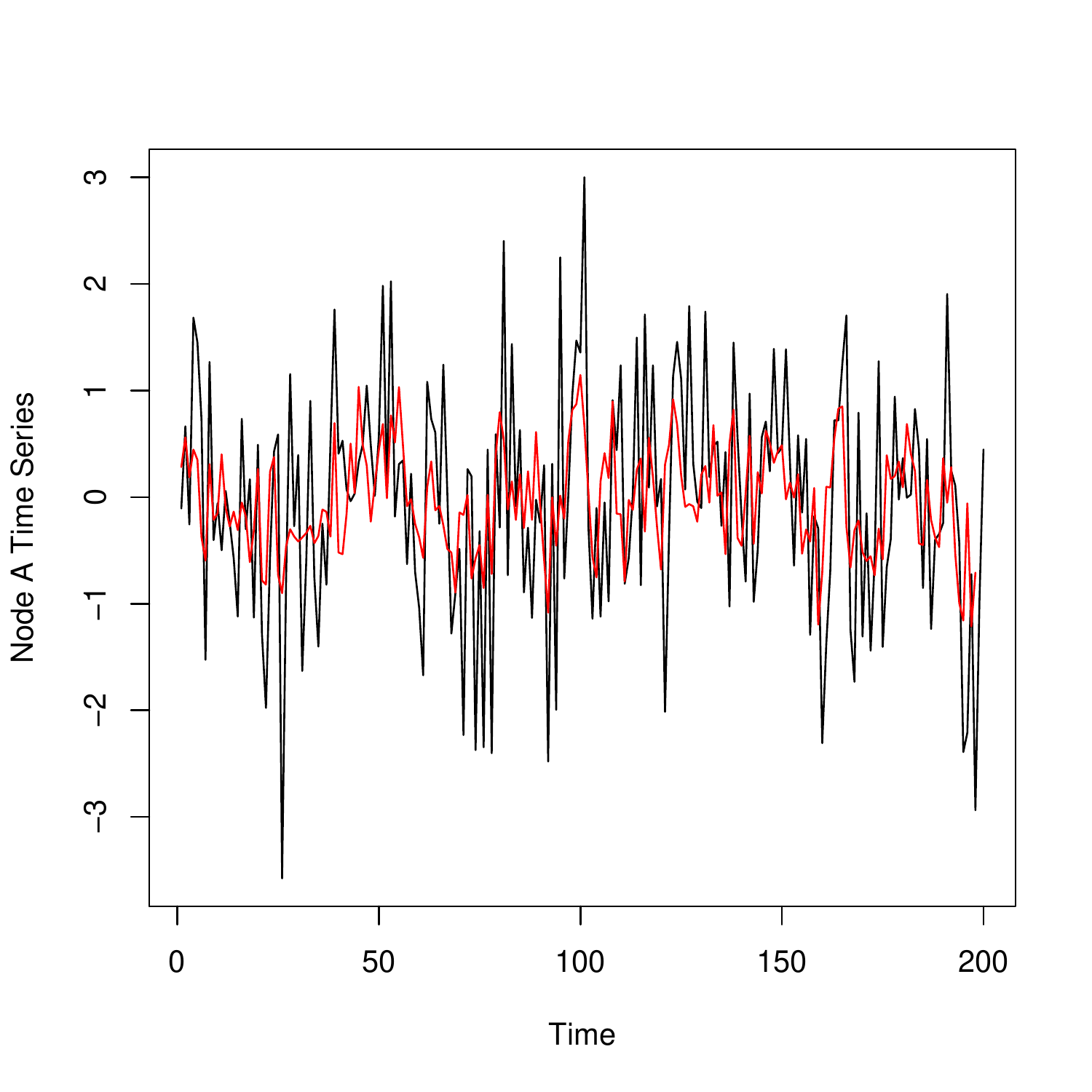}{\textwidth}{Time series of first node (black) with fitted values from `answer' model overlaid in red.}

Alternatively, we can examine the associated residuals:
\begin{Schunk}
\begin{Sinput}
R> myresiduals <- residuals(answer)[, 1]
R> layout(matrix(c(1, 2), 2, 1))
R> plot(ts(residuals(answer)[, 1]), ylab = "`answer' model residuals")
R> hist(residuals(answer)[, 1], main = "", 
+    xlab = "`answer' model residuals")
\end{Sinput}
\end{Schunk}
\myincfig{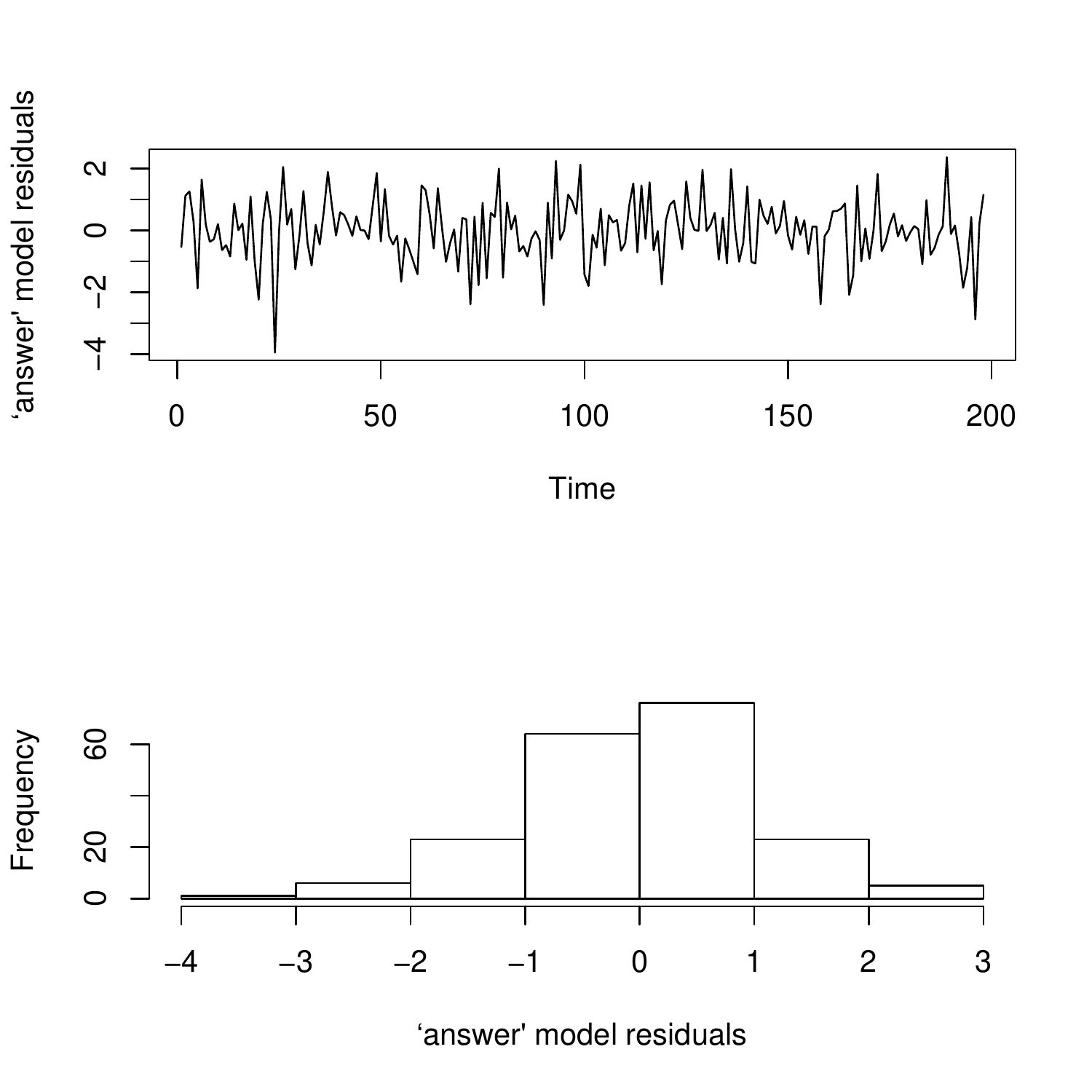}{\textwidth}{Residual plots from `answer' model fit. Top:
	Time series; Bottom: Histogram.}

By altering the input parameters in the \code{GNARfit} function, we can fit a range of different GNAR models and the reader can consult Appendix~\ref{extrasims} for further examples.

\subsection{Example: GNAR data simulation on a given network}
\label{exsimexample}
The following example demonstrates network time series simulation using the network in Figure~\ref{jss3594-pltnet}.

Model \code{(a)} is a GNAR$(1,[1])$ model with individual $\alpha$ parameters, $(\alpha_{A,1}, \alpha_{B,1}, \alpha_{C,1}, \alpha_{D,1}, \alpha_{E,1})=(0.4, 0, -0.6, 0, 0)$, and the same $\beta$ parameter throughout, $\beta_1 = 0.3$. Model \code{(b)} is a
\mbox{global-$\alpha$} GNAR$(2,[2,0])$ model with parameters $\alpha_1=0.2$, $\beta_{1,1}=0.2$, $\beta_{1,2}=0.3$ and $\alpha_2=0.3$. Both simulations are created using standard
normal noise whose standard deviation is controlled using the \code{sigma} argument.
\begin{Schunk}
\begin{Sinput}
R> set.seed(10)
R> fiveVTS2 <- GNARsim(n = 200, net = fiveNet,
+    alphaParams = list(c(0.4, 0, -0.6, 0, 0)), betaParams = list(c(0.3)))
\end{Sinput}
\end{Schunk}
By fitting an individual-alpha GNAR$(1,[1])$ model to the simulated data with the \code{fiveNet} network, we
can see that these estimated parameters are similar to the specified ones of 0.4, 0, -0.6, 0, 0 and 0.3.  This agreement does not come as a surprise given that we show theoretical consistency for parameter estimators (see Appendix~\ref{consistency}).

\begin{Schunk}
\begin{Sinput}
R> print(GNARfit(vts = fiveVTS2, net = fiveNet, alphaOrder = 1,
+    betaOrder = 1, globalalpha = FALSE))
\end{Sinput}
\begin{Soutput}
Model: 
GNAR(1,[1]) 

Call:
lm(formula = yvec ~ dmat + 0)

Coefficients:
dmatalpha1node1  dmatalpha1node2  dmatalpha1node3  dmatalpha1node4  
        0.45902          0.13133         -0.49166          0.03828  
dmatalpha1node5      dmatbeta1.1  
        0.02249          0.24848  
\end{Soutput}
\end{Schunk}

Repeating the experiment for the GNAR(2, [2, 0]) Model \code{(b)}, the estimated parameters are again similar
to the generating parameters:

\begin{Schunk}
\begin{Sinput}
R> set.seed(10)
R> fiveVTS3 <- GNARsim(n = 200, net = fiveNet,
+    alphaParams = list(rep(0.2, 5), rep(0.3, 5)),
+    betaParams = list(c(0.2, 0.3), c(0)))
R> print(GNARfit(vts = fiveVTS3, net = fiveNet, alphaOrder = 2,
+    betaOrder = c(2,0)))
\end{Sinput}
\begin{Soutput}
Model: 
GNAR(2,[2,0]) 

Call:
lm(formula = yvec ~ dmat + 0)

Coefficients:
 dmatalpha1  dmatbeta1.1  dmatbeta1.2   dmatalpha2  
     0.2537       0.1049       0.3146       0.2907  
\end{Soutput}
\end{Schunk}

Alternatively, we can use the \code{simulate} S3 method for \code{GNARfit} objects to simulate time series associated to a GNAR model, for example
\begin{Schunk}
\begin{Sinput}
R> fiveVTS4 <- simulate(GNARfit(vts = fiveVTS2, net = fiveNet, 
+    alphaOrder = 1, betaOrder = 1, globalalpha = FALSE), n = 200)
\end{Sinput}
\end{Schunk}

\subsection{Missing data and changing connection weights with GNAR models} \label{changing}
Standard multivariate time series models, including vector autoregressions (VAR), can have significant problems
in coping with certain types of missingness and imputation is often used,
see \cite{Guerrero10}, \cite{Honaker10}, \cite{Bashir16}. While in VAR modelling successful solutions have been found to cope with specific missingness scenarios, such as implemented in the \pkg{gimme} \proglang{R} package \citep{lane19:gimme}, however, if a variable has e.g., block
missing data, the coefficients corresponding that variable can be difficult to calculate, and
impossible if their partner variable is missing at cognate times. In addition, due to computational burden \pkg{gimme} is limited to modelling a single time lag. 
A key advantage
of our parsimonious GNAR model is that it models via neighbourhoods across the entire data set.
If a node is missing for a given time, then it does not contribute to the estimation of neighbourhood
parameters that the network structure suggests it should, and there are plenty of other
nodes that do contribute, generally resulting in a high number of observations to estimate
each coefficient. In GNAR models, missing data of this kind is not a problem.

The flexibility of GNAR modelling means that we can also model missing data as a changing network, or alternatively, as changing connection weights. In the situation where the overall network is considered fixed, but when observations are missing at particular nodes, the connections and weightings need altering accordingly. Again, using the graph in Figure~\ref{jss3594-pltnet}, consider the situation where node~A does not have any data recorded. Yet, we want to preserve the stage-2 connection between D and E, and the stage-3 connection between E and both B and C.
To do this, we do not redraw the graph and remove node A and its connections, instead we reweight the connections that depend on node~A. As node~A does not feature in the stage-2 or stage-3 neighbours of E, the connection weights $\omega_{E,D}, \omega_{E,B}, \omega_{E, C}$ do not change, but the connection weight $\omega_{E,A}$ drops to zero in the absence of observation from node~A. Similarly, the stage-1 neighbours of D are changed without A, so $\omega_{D,A}$ drops to zero and the other two connection weights from node~D increase accordingly; $\omega_{D,B}=\omega_{D,C} = 0.5$.

Missing data of this kind is handled automatically by the \code{GNAR} functions using
customary \code{NA} missing data values present in the \code{vts} (vector time series)
component of the overall network time series. For example, inducing some (artificial) missingness in the
\code{fiveVTS} series, we can still obtain estimates of model parameters:
\begin{Schunk}
\begin{Sinput}
R> fiveVTS0 <- fiveVTS
R> fiveVTS0[50:150, 3] <- NA
R> nafit <- GNARfit(vts = fiveVTS0, net = fiveNet, alphaOrder = 2,
+    betaOrder = c(1, 1))
R> layout(matrix(c(1, 2), 2, 1))
R> plot(ts(fitted(nafit)[, 3]), ylab = "Node C fitted values")
R> plot(ts(fitted(nafit)[, 4]), ylab = "Node D fitted values")
\end{Sinput}
\end{Schunk}
\myincfig{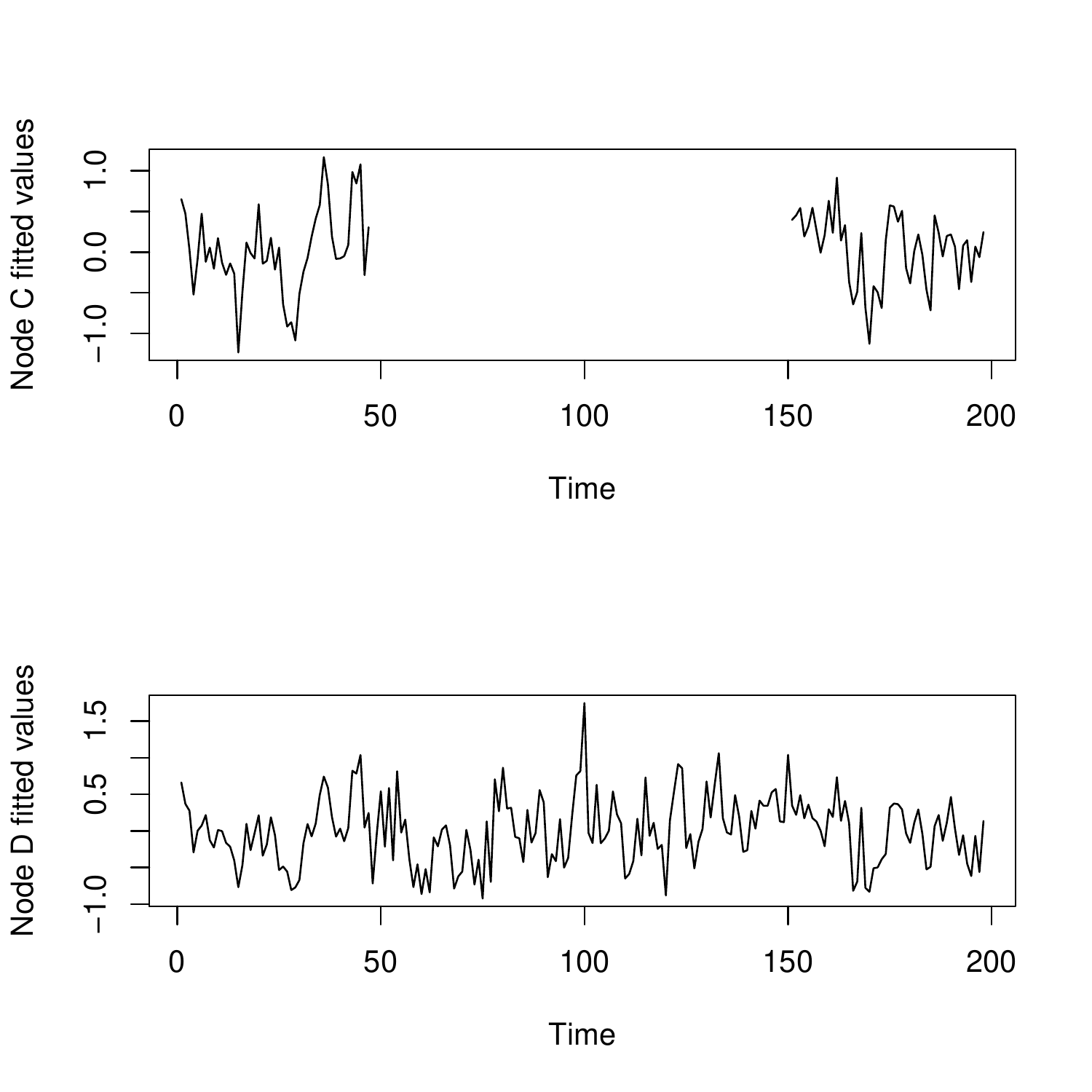}{\textwidth}{Fitted values of \mbox{global-$\alpha$} GNAR$(1,[1])$ fit to the `fiveVTS' data, with observations 50--150 removed from node C.
Fitted values: Top: Node~C; Bottom: Node~D.}
As shown in Figure~\ref{jss3594-NAs}, after removing observations from the time series at
node~C, its neighbour, node~D, still has a complete set of fitted values.
\subsection{Stationarity conditions for a GNAR process with fixed network}
\begin{theorem} \label{thstatcon}
Given an unchanging network, $\mathcal{G}$, a sufficient condition for the GNAR model (\ref{NAR1}) to be stationary is
\begin{equation}\label{statcon}
\sum_{j=1}^p \left( |\alpha_{i,j}| + \sum\limits_{c=1}^C \sum\limits_{r=1}^{s_j}|\beta_{j,r,c}| \right)<1 \quad \forall i \in 1,...,N.
\end{equation}
\end{theorem}
\noindent The proof of Theorem~\ref{thstatcon} can be found in Appendix~\ref{appstatcon}.

For the global-$\alpha$ model this condition reduces to
\begin{equation}
    {\sum_{j=1}^p \left( |\alpha_{j}| + \sum\limits_{c=1}^C  \sum\limits_{r=1}^{s_j}|\beta_{j,r,c}| \right)<1 }.
\end{equation}
We can explore these conditions using the~\code{GNARsim} function.
The following example uses parameters whose absolute value sums to greater than one and
then we calculate the mean over successive time periods. The mean increases rapidly
indicating nonstationarity.
\begin{Schunk}
\begin{Sinput}
R> set.seed(10)
R> fiveVTS4 <- GNARsim(n = 200, net = fiveNet,
+    alphaParams = list(rep(0.2, 5)), betaParams = list(c(0.85)))
R> c(mean(fiveVTS4[1:50, ]), mean(fiveVTS4[51:100, ]),
+    mean(fiveVTS4[101:150, ]), mean(fiveVTS4[151:200, ]))
\end{Sinput}
\begin{Soutput}
[1]    -120.511   -1370.216  -15725.884 -180319.140
\end{Soutput}
\end{Schunk}
\subsection{Benefits of our model and comparisons to others}
Conditioned on a given network fixed in time and with a known (time-dependent) weight- and neighbourhood structure, the
GNAR model can be mathematically formulated as a specific restricted VAR model, where the restrictions are imposed by the network and thus impact model parametrisation,
as mathematically encoded by equation~\eqref{NAR1}. This is explored in more depth in Appendix~\ref{consistency} and contrasts with a VAR model where any restrictions
can only be imposed on the parameters themselves.


An unrestricted VAR model with dimension $n$ has $\mathcal{O}(n^2)$ parameters, whereas a GNAR model with known network (usually) has $\mathcal{O}(n)$ parameters, and a global-$\alpha$ GNAR model can have $\mathcal{O}(1)$ parameters.
The large, and rapidly increasing, number of parameters in VAR often make it a challenging
model to fit and non-problem-specific mathematical constraints are often used to mitigate those
challenges. Further, the large number of VAR parameters usually mean that it fits multivariate
time series well, but then performs poorly in out-of-sample prediction.
An example of this is shown in Section~\ref{gdp}.

Our model has similarities with the network autoregression
introduced by \cite{Zhu2017}, motivated by social networks.

In our notation, the \cite{Zhu2017} model can be written as a special case as
\begin{equation}
    X_{i,t} = \beta_0 + Z_i^\top\gamma + \sum_{j=1}^p \left( \alpha_j X_{i,t-j} + \beta_j \sum_{q \in \mathcal{N}^{(1)}(i)} \omega_i X_{q,t-j} \right) + u_{i,t},
\end{equation}
where $\beta_0$ is a global intercept term, $Z_i$ is a vector of node-specific covariates with corresponding parameters $\gamma$, $\omega_i$ is the reciprocal of the out-degree of node $i$, and the innovations are independent and identically distributed, with zero mean, such that $\var(u_{i,t})=\sigma^2$. Hence,  the \cite{Zhu2017} model without intercept and node-specific covariates is a special case of our GNAR model, with $\max\limits_{j \in \{1,...,p\}} s_j = 1$, i.e., dependencies limited to stage-1 immediate neighbours, and un-weighted edges.


Our model is designed to deal with a time-varying network, and our $\beta_{j, r, c}$ parameters can include general edge-based covariate information.
A further important advantage is that our GNAR model in Section~\ref{NARmod} can express dependence on stage-$r$ neighbour sets for any $r$.

An earlier model with similarities to the generic network autoregression
is the Dynamic Bayesian Network (DBN) model considered in \cite{Spencer2015}. Their model can be written as
\begin{equation}
    X_{i,t} = \beta_{0,i} + \sum\limits_{q \in \mathcal{N}^{(1)}(i)} \beta_{i,q} X_{q,t-1} + u_{i,t},
\end{equation}
where $\beta_{0,i}$ is a node-specific intercept term, the other $\beta$ parameters describe the network autoregression, and $u_{i,t} \sim N(0, \sigma_i^2)$.
The DBN model is also a constrained VAR model, but with no univariate autoregression terms,
and the network autoregression only includes the stage-1 neighbours.
Unlike our model and the \cite{Zhu2017} model, there are no restrictions on the parameters other than parameters only being present when there is an edge between two nodes. The \cite{Spencer2015} framework does not allow for a range of networks, as their underlying network  is assumed to be a Directed Acyclic Graph. With these assumptions, the network and parameters are inferred by considering potential predictors for each node in turn. A key difference between our model and the \cite{Spencer2015} model is that we  assume that the behaviour of connected nodes is the same throughout the network, whereas the DBN model allows for different $\beta$ parameters for different connections, including allowing a change of sign.

The benefits of the GNAR model compared to these, and other models, include the ability to deal with a time-changing network, missing observations,
and using network information to reduce the number of parameters.
As detailed in Section~\ref{changing}, we can incorporate missing data information with the GNAR model by allowing the connection weights to change. Allowing for a changing network structure enables us to model new nodes being added to the system, or connections between nodes changing over time. Adding autoregressive parameters to neighbours with stage greater than one results in our model being able to capture more network relationships than just those of immediate neighbours.


\section{Estimation} \label{Implementation}
In modelling terms, our GNAR model is a linear model and we employ standard techniques such as least squares estimation to fit them and to provide
statistically consistent estimators, as verified in Appendix~\ref{consistency}. An important practical consideration for fitting GNAR models is the choice of
model order.  Specifically, how do we  select $p$ and $\mathbf{s}$?

\subsection{Order selection} \label{BIC}
We use the Bayesian information criterion (BIC) proposed by~\cite{Schwarz1978}
to select the GNAR model order.
Under the assumption of a constant network, and that the innovations are independent
and identically distributed white noise with bounded fourth moments, this criterion is consistent,
as shown in~\cite{Lutkepohl2005}. The BIC allows us to select both the lag and neighbourhood orders simultaneously by selecting the model with smallest BIC from a set of candidates.

For a  general candidate GNAR$(p, [\mathbf{s}])$ model with $N$ nodes,
the BIC is given by
\begin{equation} \label{BICeq}
    \operatorname{BIC}(p,\mathbf{s}) = \ln | \hat{\varSigma}_{p,\mathbf{s}} | + T^{-1} M\ln(T) ,
\end{equation}
where ${\hat{\varSigma}_{p,\mathbf{s}} = T^{-1} \hat{\mathit{U}}'\hat{\mathit{U}}}$, $\hat{\mathit{U}}$ is the residual matrix from  the
NAR$(p, [\mathbf{s}])$ fit, and $M$ is the number of parameters. In the general case $M=Np +C \sum_{j=1}^p s_j$, and in the global-$\alpha$ model $M=p+C\sum_{j=1}^p s_j$. The covariance matrix estimate, $\hat{\varSigma}_{p,\mathbf{s}}$, is also the maximum likelihood estimator of the innovation covariance matrix under the assumption of Gaussian innovations.

\pkg{GNAR} enables us to easily compute the BIC for any model by using
the \code{BIC} method for \code{GNARfit} objects. For example, on the default model fitted by
\code{GNARfit}, and an alternative model that additionally includes second-order
neighbours at the first lag into the model, we can compare their BICs by
\begin{Schunk}
\begin{Sinput}
R> BIC(GNARfit())
\end{Sinput}
\begin{Soutput}
[1] -0.003953124
\end{Soutput}
\begin{Sinput}
R> BIC(GNARfit(betaOrder = c(2, 1)))
\end{Sinput}
\begin{Soutput}
[1] 0.02251406
\end{Soutput}
\end{Schunk}

Whilst we focus on the BIC for model selection for the remainder of this article, the \pkg{GNAR} package also
include functionality for the Akaike information criterion (AIC) proposed by \cite{akaike73:information} as
\begin{equation} \label{AICeq}
    \operatorname{AIC}(p,\mathbf{s}) = \ln | \hat{\varSigma}_{p,\mathbf{s}} | + 2 T^{-1} M,
\end{equation}
where ${\hat{\varSigma}_{p,\mathbf{s}}}$ is as defined in equation~\eqref{BICeq} and $M$ is again the number of
model parameters.  Similar to above, the AIC can be obtained by using the code
\begin{Schunk}
\begin{Sinput}
R> AIC(GNARfit())
\end{Sinput}
\begin{Soutput}
[1] -0.06991947
\end{Soutput}
\begin{Sinput}
R> AIC(GNARfit(betaOrder = c(2, 1)))
\end{Sinput}
\begin{Soutput}
[1] -0.05994387
\end{Soutput}
\end{Schunk}
Similar to the BIC, the model with the lowest AIC is preferred.  Note that the likelihood of the data associated to the model fit can also be obtained using e.g., \code{logLik(GNARfit())}.

Various models can be tried to obtain a good fit whilst, naturally, attending to
the usual aspects of good model fitting, such as residual checks. A thorough simulation study that displays the numerical performance of our proposed method appears in Section 4.5 of \cite{LeemingPhD}.

\subsection{Model selection on a wind network time series}
\label{wind}
\pkg{GNAR} incorporates the data suite \code{vswind} that contains a number of
\proglang{R} objects pertaining to 721 wind speeds taken at each of 102 weather
stations in England and Wales. The suite contains the vector time
series \code{vswindts}, the associated network \code{vswindnet},
a character vector of the weather station location names in \code{vswindnames}
and coordinates of the stations in the two column matrix \code{vswindcoords}.
The data originate from the UK Met Office site
\url{http://wow.metoffice.gov.uk} and full details can be found in
the \code{vswind}
help file in the \pkg{GNAR} package.
Figure~\ref{jss3594-windnetplot} shows a picture of the meteorological station network with distances
created by
\begin{Schunk}
\begin{Sinput}
R> oldpar <- par(cex = 0.75)
R> windnetplot()
R> par(oldpar)
\end{Sinput}
\end{Schunk}
\myincfig{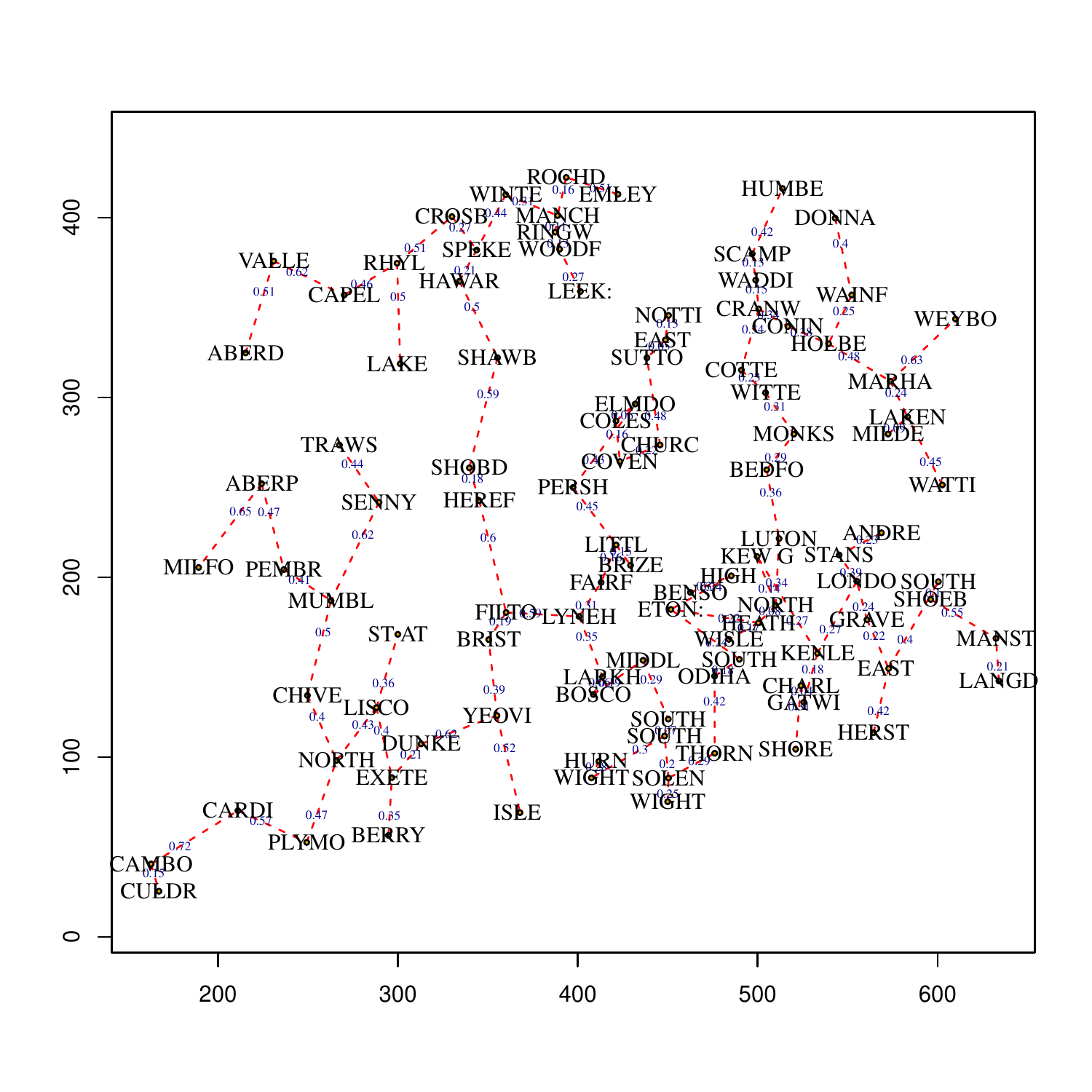}{\textwidth}{Plot of the wind speed network. Blue numbers
	are relative distances between sites; labels are the site name.}
We investigate fitting a network time series model. We first fit a simple GNAR$(1, [0])$ model using
a single $\alpha$, followed by an equivalent model with potentially individually distinct $\alpha$s
\begin{Schunk}
\begin{Sinput}
R> BIC(GNARfit(vts = vswindts, net = vswindnet, alphaOrder = 1,
+    betaOrder = 0))
\end{Sinput}
\begin{Soutput}
[1] -233.3848
\end{Soutput}
\begin{Sinput}
R> BIC(GNARfit(vts = vswindts, net = vswindnet, alphaOrder = 1,
+    betaOrder = 0, globalalpha = FALSE))
\end{Sinput}
\begin{Soutput}
[1] -233.1697
\end{Soutput}
\end{Schunk}
Interestingly, the model with the single $\alpha$ gives the better fit, as judged by BIC.
The single $\alpha$ model with \code{alphaOrder = 2} and \code{betaOrder = c(0, 0)} gives a lower BIC of
$-243$, so we investigate this next.  Note that this model also gives the lowest AIC score. In particular, we explore a set of GNAR$(2, [b1, b2])$ models
with $b1$, $b2$ ranging from zero to 14 using the following code:
\begin{Schunk}
\begin{Sinput}
R> BIC.Alpha2.Beta <- matrix(0, ncol = 15, nrow = 15)
R> for(b1 in 0:14)
+    for(b2 in 0:14)
+      BIC.Alpha2.Beta[b1 + 1, b2 + 1] <- BIC(GNARfit(vts = vswindts,
+        net = vswindnet, alphaOrder = 2, betaOrder = c(b1, b2)))
R> contour(0:14, 0:14, log(251 + BIC.Alpha2.Beta),
+    xlab = "Lag 1 Neighbour Order", ylab = "Lag 2 Neighbour Order")
\end{Sinput}
\end{Schunk}
\myincfig{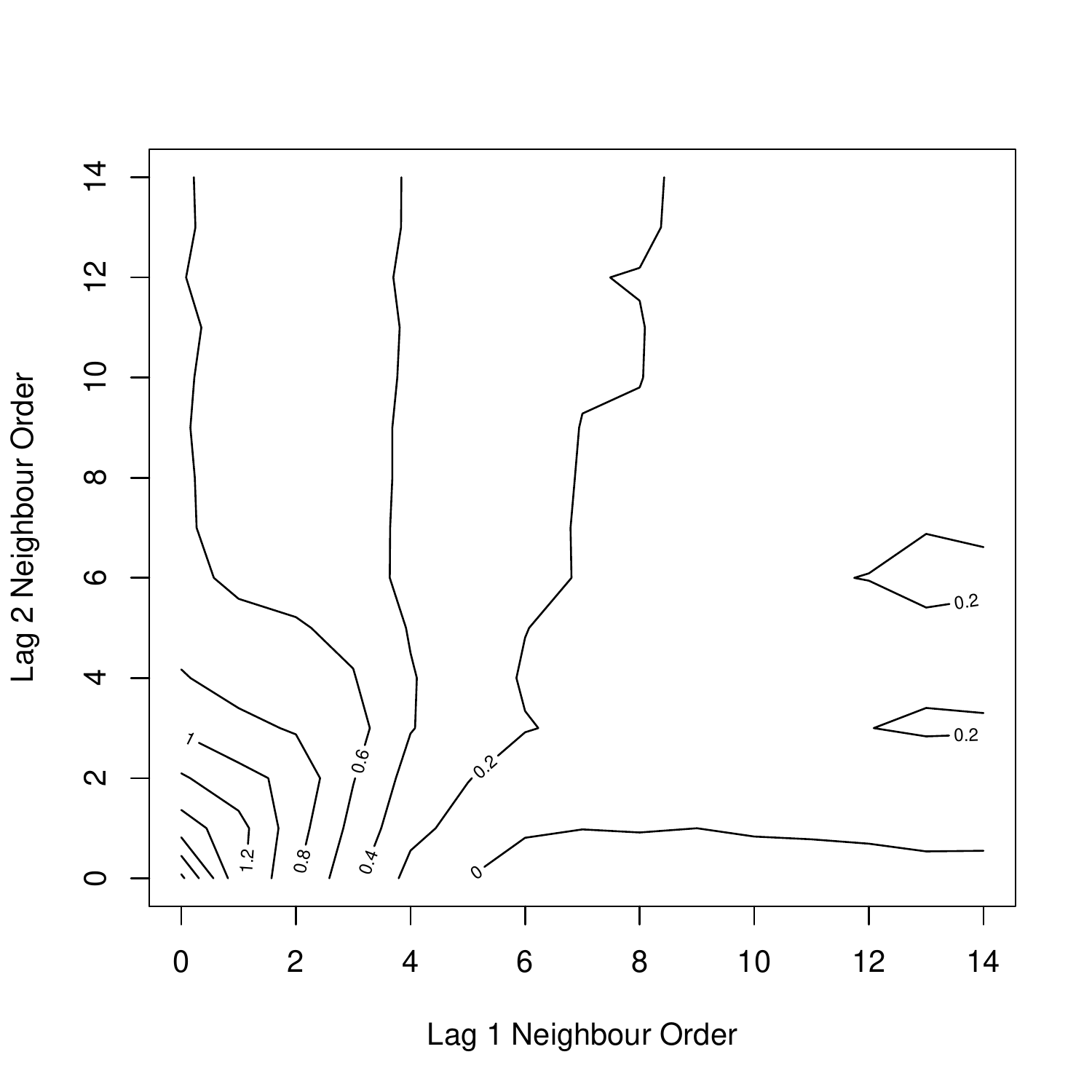}{\textwidth}{Contour plot of BIC values for the two-lag autoregressive model
incorporating \mbox{$b1$-stage} and \mbox{$b2$-stage} neighbours at time lags one and two.
Values shown are $\log(251 + \operatorname{BIC})$ to display clearer contours.}
The results of the BIC evaluation for incorporating different and deeper neighbour sets, at lags one
and two, are
shown in the contour plot in Figure~\ref{jss3594-windcontour}. The minimum value of the BIC occurs in the bottom-right
part of the plot, where it seems incorporating five or sixth-stage neighbours for the first time lag
is sufficient to achieve the minimum BIC, and incorporating further lag one stages does not reduce the BIC. Moreover,
increasing the lag two neighbour sets beyond first stage neighbours would appear to increase the BIC for those
lag one neighbour stages greater than five (the horizontal contour at $0$ in the bottom right hand corner of
the plot). A fit of a possible model is
\begin{Schunk}
\begin{Sinput}
R> goodmod <- GNARfit(vts = vswindts, net = vswindnet, alphaOrder = 2,
+    betaOrder = c(5, 1))
R> goodmod
\end{Sinput}
\begin{Soutput}
Model: 
GNAR(2,[5,1]) 

Call:
lm(formula = yvec ~ dmat + 0)

Coefficients:
 dmatalpha1  dmatbeta1.1  dmatbeta1.2  dmatbeta1.3  dmatbeta1.4  
    0.56911      0.10932      0.03680      0.02332      0.02937  
dmatbeta1.5   dmatalpha2  dmatbeta2.1  
    0.04709      0.23424     -0.04872  
\end{Soutput}
\end{Schunk}
We investigated  models with \code{alphaOrder} equal to two, three, four and five, but with no
neighbours. As judged by BIC, \code{alphaOrder = 3}
gives the best model. We could extend the example above to investigate differing stages of neighbours at
time lags one, two and three. However, a more comprehensive BIC investigation would examine
all combinations of neighbour sets over a large number of time lags. This would be feasible, but computationally
intensive for a single CPU machine, but could be coarse-grain parallelized.
Further analysis would proceed with model diagnostic checking and further modelling
as necessary.


\subsection{Constructing a network to aid prediction}\label{construction}
Whilst some multivariate time series have actual, and sometimes obvious, networks associated with them, our methodology can be useful for series without a clear or
supplied network. We propose a network construction method that uses prediction error, but note here that our scope is not to estimate an underlying network, but
merely to find a structure that is useful in the task of prediction. Here, we use a prediction error measure, understood as the sum of squared differences between the
observations and the estimates: $\sum_{i=1}^N (X_{i,t} - \hat{X}_{i,t})^2$.

The \code{predict} S3 method for GNAR models takes an input \code{GNARfit} model object and from this predicts the nodal time series at the next timepoint, similar to the S3 method for the \code{Arima} class.  This allows for a `ex-sample' prediction evaluation. The \code{predict} function outputs the prediction as a vector. For example, to predict the series at the last timepoint
\begin{Schunk}
\begin{Sinput}
R> prediction <- predict(GNARfit(vts = fiveVTS[1:199,], net = fiveNet,
+    alphaOrder = 2, betaOrder = c(1, 1)))
R> prediction
\end{Sinput}
\begin{Soutput}
Time Series:
Start = 1 
End = 1 
Frequency = 1 
    Series 1  Series 2  Series 3  Series 4   Series 5
1 -0.6427718 0.2060671 0.2525534 0.1228404 -0.8231921
\end{Soutput}
\end{Schunk}

For a small-dimensional multivariate series, any and all
potential un-weighted networks can be constructed and the corresponding prediction errors compared using the \code{predict} method. Next, we consider the larger data setting where it is computationally infeasible to investigate all possible networks.
Erd\H{o}s-R\'{e}nyi random graphs can be generated with $N$ nodes, and a fixed probability of including each edge between these nodes, see Chapter~11 of \cite{Grimmett2010} for further details. The probability parameter controls the overall sparsity of the graph. Many random graphs of this type can be created, and then our GNAR model can be used for within-sample prediction. The prediction error can then be used to identify networks that aid prediction. We give an example of
this process in the next section.


\section{OECD  GDP: Network structure aids prediction} 
\label{gdp}
We obtained the annual gross domestic product (GDP) growth rate time series for
35 countries from the OECD website\footnote{OECD (2018), Quarterly GDP (indicator). doi: 10.1787/b86d1fc8-en (Accessed on 29 January 2018)}. The  series
covers the years 1961--2013, but not all countries are included from the start.
The values are annual growth rates expressed as a percentage change compared to the previous year. We differenced the time series for each country to remove
the gross trend.

We use the first $T=52$ time points and designate each of the 35 countries as nodes to investigate the potential of modelling this time series using a network. In this data set 20.8\% (379 out of 1820) of the observations were missing due to some nodes not being included from the start.
We model this by changing the network connection weights as described in
Section~\ref{changing}. In this example, we do not use covariate information, so $C=1$.
The pattern of missing data along with the time series values is shown
graphically in Figure~\ref{jss3594-gdpheat}, produced by the following code.
\begin{Schunk}
\begin{Sinput}
R> library("fields")
R> layout(matrix(c(1, 2), nrow = 1, ncol = 2), widths = c(4.5, 1))
R> image(t(apply(gdpVTS, 1, rev)), xaxt = "n", yaxt = "n",
+  	col = gray.colors(14), xlab = "Year", ylab = "Country")
R> axis(side = 1, at = seq(from = 0, to = 1, length = 52), labels = FALSE,
+    col.ticks = "grey")
R> axis(side = 1, at = seq(from = 0, to = 1, length = 52)[5*(1:11)],
+    labels = (1:52)[5*(1:11)])
R> axis(side = 2, at = seq(from = 1, to = 0, length = 35),
+    labels = colnames(gdpVTS), las = 1, cex = 0.8)
R> layout(matrix(1))
R> image.plot(zlim = range(gdpVTS, na.rm = TRUE), legend.only = TRUE,
+    col = gray.colors(14))
\end{Sinput}
\end{Schunk}
\myincfig{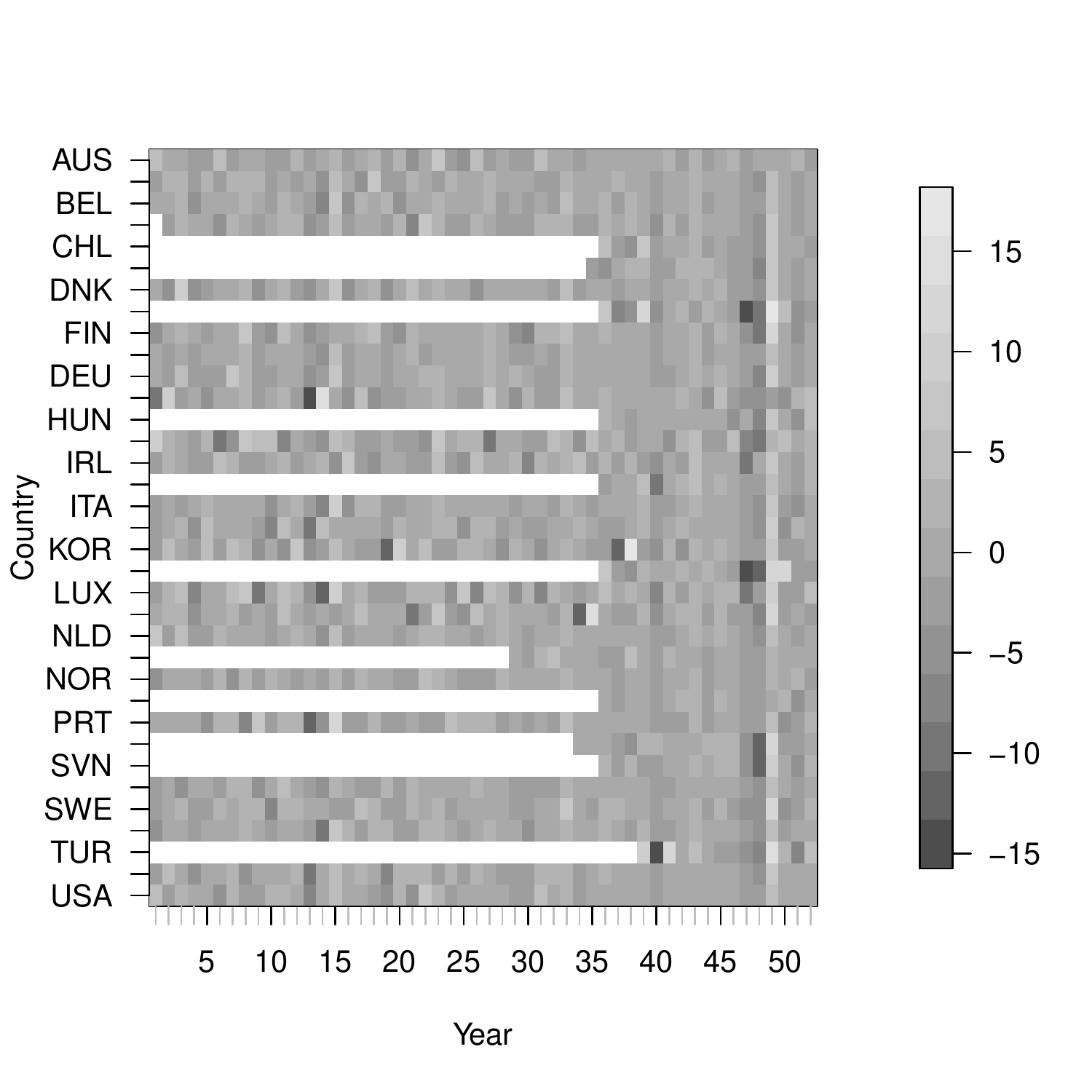}{\textwidth}{Heat plot (greyscale) of the differenced time series, where the initial white space indicates missing time series observations.}

\subsection{Finding a network to aid prediction}
\label{sec:findnetaidp}
This section considers the case where we  observe data up to $t=51$,
and then wish to predict the values for each node at $t=52$. We begin by exploring `within-sample' prediction at $t=51$, and identify a good network for prediction. We use randomly generated Erd\H{o}s-R\'{e}nyi graphs using the \pkg{GNAR} function \code{seedToNet}.
To demonstrate this, the \pkg{GNAR} package contains the \code{gdp} data and a
set of seed values, \code{seed.nos} so that the random graphs can be reproduced for use with the time series object \code{gdpVTS}
here.
\begin{Schunk}
\begin{Sinput}
R> net1 <- seedToNet(seed.no = seed.nos[1], nnodes = 35, graph.prob = 0.15)
R> net2 <- seedToNet(seed.no = seed.nos[2], nnodes = 35, graph.prob = 0.15)
R> layout(matrix(c(2, 1), 1, 2))
R> par(mar=c(0,1,0,1))
R> plot(net1, vertex.label = colnames(gdpVTS), vertex.size = 0)
R> plot(net2, vertex.label = colnames(gdpVTS), vertex.size = 0)
\end{Sinput}
\end{Schunk}
\myincfig{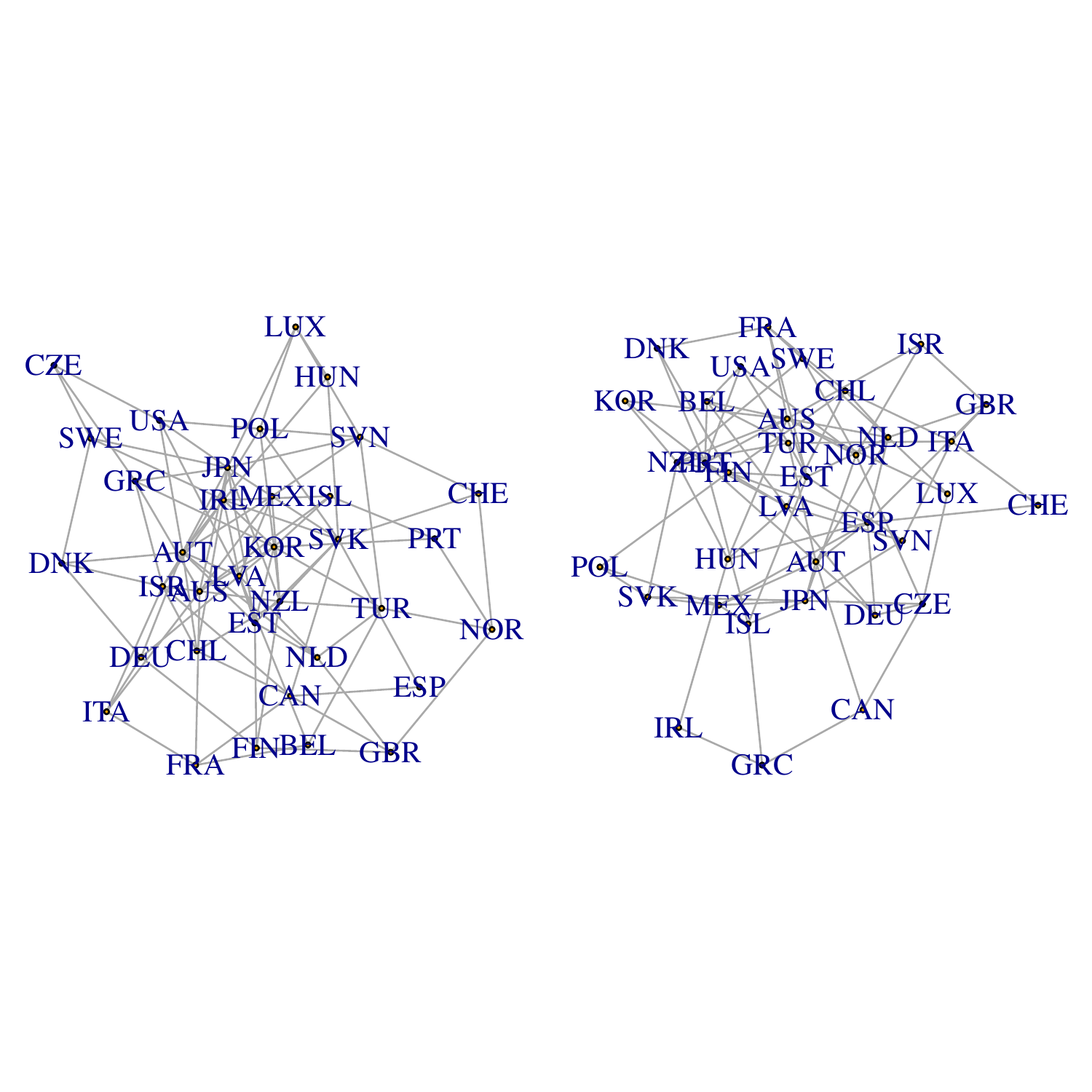}{\textwidth}{Erd\H{o}s-R\'{e}nyi random graphs constructed from the first two elements of the \code{seed.nos} variable with 35 nodes and connection probability 0.15.}
Figure~\ref{jss3594-seedToNet1} shows two of these random graphs.

As well as investigating which network works best for prediction, we also need to identify the number of parameters in the GNAR model. Initial analysis of the autocorrelation function at each node indicated that a second-order autoregressive component should be sufficient, so GNAR models with orders up to $p=2$ were tested, and we included at most two neighbour sets at each time lag. The GNAR models are: GNAR$(1,[0])$, GNAR$(1,[1])$, GNAR$(2,[0,0])$, GNAR$(2,[1,0])$, GNAR$(2,[1,1])$,
GNAR$(2,[2,0])$, GNAR$(2,[2,1])$, and GNAR$(2,[2,2])$,
each fitted as individual-$\alpha$ and global-$\alpha$ GNAR models,
giving sixteen models in total.

\begin{figure}
    \centering
    \includegraphics[angle=270, scale=1.2]{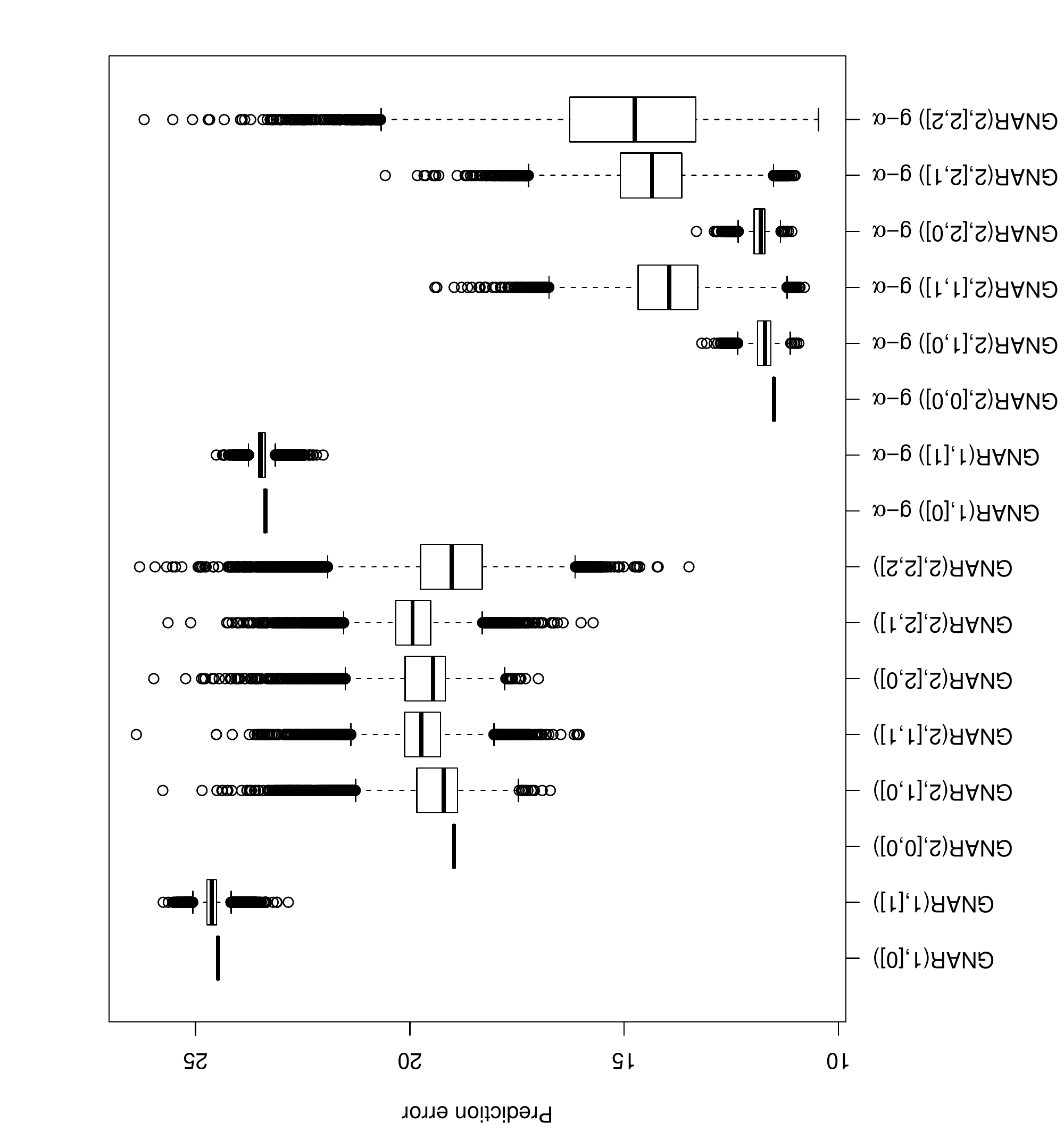}
    \caption{Prediction error boxplots at $t=51$ over 10,000 randomly generated networks using \code{seed.nos} and different GNAR models, where `g-$\alpha$' indicates a global-$\alpha$ GNAR model.}
    \label{fig:0209boxres}
\end{figure}

For the GDP example, we simulate 10,000 random un-directed networks, each with connection probability 0.15, and predict using the GNAR model with the orders above. Hence, this example requires significant computation time (about 90 minutes on
a desktop PC),
so only a segment of the analysis is included in the code below. For computational reasons, we first divide through by the standard deviation at each node so that we can model the residuals as having equal variances at each node. The function \code{seedSim} outputs the sum of squared differences between the prediction and original values, and we use this as our measure of prediction accuracy.
\begin{Schunk}
\begin{Sinput}
R> gdpVTSn <- apply(gdpVTS, 2, function(x){x / sd(x[1:50], na.rm = TRUE)})
R> alphas <- c(rep(1, 2), rep(2, 6))
R> betas <- list(c(0), c(1), c(0, 0), c(1, 0), c(1, 1), c(2, 0), c(2, 1),
+    c(2, 2))
R> seedSim <- function(seedNo, modelNo, globalalpha){
+    net1 <- seedToNet(seed.no = seedNo, nnodes = 35, graph.prob = 0.15)
+    gdpPred <- predict(GNARfit(vts = gdpVTSn[1:50, ], net = net1,
+      alphaOrder = alphas[modelNo], betaOrder = betas[[modelNo]],
+      globalalpha = globalalpha))
+    return(sum((gdpPred - gdpVTSn[51, ])^2))
+  }
R> seedSim(seedNo = seed.nos[1], modelNo = 1, globalalpha = TRUE)
\end{Sinput}
\begin{Soutput}
[1] 23.36913
\end{Soutput}
\begin{Sinput}
R> seedSim(seed.nos[1], modelNo = 3, globalalpha = TRUE)
\end{Sinput}
\begin{Soutput}
[1] 11.50739
\end{Soutput}
\begin{Sinput}
R> seedSim(seed.nos[1], modelNo = 3, globalalpha = FALSE)
\end{Sinput}
\begin{Soutput}
[1] 18.96766
\end{Soutput}
\end{Schunk}
Prediction error boxplots over simulations from
all sixteen models and 10,000 random networks are shown in Figure~\ref{fig:0209boxres}
(accompanying code not shown due to significant computation time).
The global-$\alpha$ model resulted in lower prediction error in general, so we use this version of the GNAR model. For GNAR$(1,[0])$ and GNAR$(2,[0,0])$, the first and third model
in Figure~\ref{fig:0209boxres} the ``boxplots'' are short horizontal lines as the
results for each graph are identical, as no neighbour parameters are fitted.
As the other global-$\alpha$ models are nested within it, we select the randomly generated graph that minimises the prediction error for global-$\alpha$ GNAR$(2,[2,2])$; this turns out
to be the network generated from \code{seed.nos[921]}.
\begin{Schunk}
\begin{Sinput}
R> net921 <- seedToNet(seed.no = seed.nos[921], nnodes = 35,
+    graph.prob = 0.15)
R> layout(matrix(c(1), 1, 1))
R> plot(net921, vertex.label = colnames(gdpVTS), vertex.size = 0)
\end{Sinput}
\end{Schunk}
\myincfig{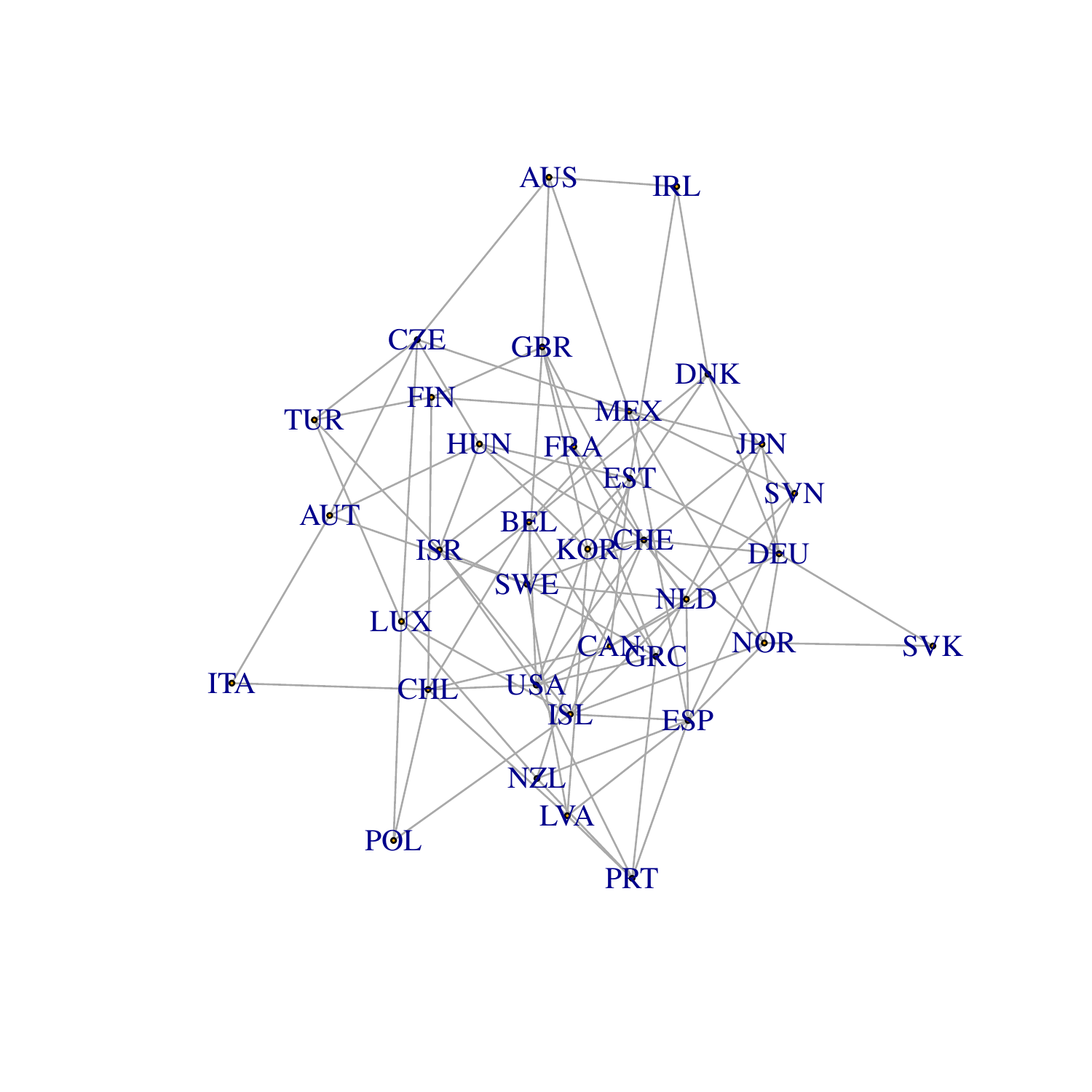}{\textwidth}{Randomly generated un-weighted and un-directed graph over the OECD countries that minimises the prediction error at $t=51$ using GNAR$(2,[2,2])$.}
The network generated from \code{seed.nos[921]} is plotted in Figure~\ref{jss3594-seed921net},
where all countries have at least two neighbours, with 97 edges in total.
This ``921''
network was constructed with GDP prediction in mind, so we would not necessarily
expect any interpretable structure in our found network (and presumably, there
were other networks with not too dissimilar predictive power). However,
the USA, Mexico and Canada are extremely well-connected with eight, eight
and six edges, respectively. Sweden and Chile are also well-connected,
with eight and seven edges, respectively. This might seem surprising, but,
e.g., the McKinsey Global Institute MGI Connectedness Index, see~\cite{McKinsey2016},
ranks Sweden and Chile 18th and 45th respectively out of 139 countries, and
each country is most connected within their regional bloc (Nordic and South America,
respectively).
Each of these edges, or subgraphs of the ``921''
network could be tested to find a sparser network with a similar predictive performance, but we continue with the full chosen network here.

Using this network, we can select the best GNAR order using the BIC.
\begin{Schunk}
\begin{Sinput}
R> res <- rep(NA, 8)
R> for(i in 1:8){
+    res[i] <- BIC(GNARfit(gdpVTSn[1:50, ],
+      net = seedToNet(seed.nos[921], nnodes = 35, graph.prob = 0.15),
+      alphaOrder = alphas[i], betaOrder = betas[[i]]))
+  }
R> order(res)
\end{Sinput}
\begin{Soutput}
[1] 6 3 4 7 8 5 1 2
\end{Soutput}
\begin{Sinput}
R> sort(res)
\end{Sinput}
\begin{Soutput}
[1] -64.44811 -64.32155 -64.18751 -64.12683 -64.09656 -63.86919
[7] -60.67858 -60.54207
\end{Soutput}
\end{Schunk}
The model that minimised BIC in this case was the sixth model, GNAR$(2,[2,0])$,
a model with two autoregressive parameters and network regression parameters on the first two neighbour sets at time lag one.

\subsection{Results and comparisons}
\label{sec:results}
We use the previous section's model to predict the values at $t=52$
and compare its prediction errors to those found using standard
AR and VAR models.
The GNAR predictions are found by fitting a GNAR$(2,[2,0])$ model with the chosen network (corresponding to \code{seed.nos[921]}) to data up to $t=51$, and then predicting  values at $t=52$.  We first normalise the series, and then compute the total squared error from the model fit.
\begin{Schunk}
\begin{Sinput}
R> gdpVTSn2 <- apply(gdpVTS, 2, function(x){x / sd(x[1:51], na.rm = TRUE)})
R> gdpFit <- GNARfit(gdpVTSn2[1:51,], net = net921, alphaOrder = 2,
+    betaOrder = c(2, 0))
R> summary(gdpFit)
\end{Sinput}
\begin{Soutput}
Call:
lm(formula = yvec2 ~ dmat2 + 0)

Residuals:
    Min      1Q  Median      3Q     Max 
-3.4806 -0.5491 -0.0121  0.5013  3.1208 

Coefficients:
             Estimate Std. Error t value Pr(>|t|)    
dmat2alpha1  -0.41693    0.03154 -13.221  < 2e-16 ***
dmat2beta1.1 -0.12662    0.05464  -2.317   0.0206 *  
dmat2beta1.2  0.28044    0.06233   4.500  7.4e-06 ***
dmat2alpha2  -0.33282    0.02548 -13.064  < 2e-16 ***
---
Signif. codes:  0 '***' 0.001 '**' 0.01 '*' 0.05 '.' 0.1 ' ' 1

Residual standard error: 0.8926 on 1332 degrees of freedom
  (23 observations deleted due to missingness)
Multiple R-squared:  0.1859,	Adjusted R-squared:  0.1834 
F-statistic: 76.02 on 4 and 1332 DF,  p-value: < 2.2e-16

GNAR BIC: -62.86003
\end{Soutput}
\begin{Sinput}
R> sum((predict(gdpFit) - gdpVTSn2[52, ])^2)
\end{Sinput}
\begin{Soutput}
[1] 5.737203
\end{Soutput}
\end{Schunk}
The fitted parameters of this GNAR model were $\hat{\alpha}_1 \simeq -0.42, \hat{\beta}_{1,1} \simeq -0.13, \hat{\beta}_{1,2} \simeq 0.28,$ and $ \hat{\alpha}_2 \simeq -0.33$.

We compared our methods with results from fitting an AR model individually
to each node using the  \code{forecast.ar()} and \code{auto.arima()} functions from version~8.0 of the CRAN \pkg{forecast} package \citep{hyndman17:forecast},
for further details see \cite{Hyndman2008}.
Due to our autocorrelation analysis from Section~\ref{sec:findnetaidp}
we set the maximum AR order for each of the 35 individual models
to be $p=2$. Conditional on this, the actual order selected was chosen using
the BIC.
\begin{Schunk}
\begin{Sinput}
R> library("forecast")
R> arforecast <- apply(gdpVTSn2[1:51, ], 2, function(x){
+    forecast(auto.arima(x[!is.na(x)], d = 0, D = 0, max.p = 2, max.q = 0,
+      max.P = 0, max.Q = 0, stationary = TRUE, seasonal = FALSE, ic = "bic",
+      allowmean = FALSE, allowdrift = FALSE, trace = FALSE), h = 1)$mean
+    })
R> sum((arforecast - gdpVTSn2[52, ])^2)
\end{Sinput}
\begin{Soutput}
[1] 8.065491
\end{Soutput}
\end{Schunk}
Our VAR comparison was calculated using version~1.5--2 of the
CRAN package \pkg{vars}, \cite{Pfaff2008}. The missing values at the beginning of the series cannot be handled with current software, so are set to zero. The number of parameters in a zero-mean VAR($p$) model is
of order $pN^2$.
In this particular example, the dimension of the observation data matrix is
$T\times N$, with $T < 2N$, so only a first-order VAR can be fitted.
We fit the model using the \code{VAR} function and then use
the \code{restrict} function to reduce dimensionality further, by setting to zero
any coefficient whose associated absolute \mbox{$t$-statistic} value is less than two.
\begin{Schunk}
\begin{Sinput}
R> library("vars")
R> gdpVTSn2.0 <- gdpVTSn2
R> gdpVTSn2.0[is.na(gdpVTSn2.0)] <- 0
R> varforecast <- predict(restrict(VAR(gdpVTSn2.0[1:51, ], p = 1,
+    type = "none")), n.ahead = 1)
\end{Sinput}
\end{Schunk}

This results in forecast vectors for each node, so we extract the point forecast (the first element of the
forecast vectors) and compute the prediction error as follows
\begin{Schunk}
\begin{Sinput}
R> getfcst <- function(x){return(x[1])}
R> varforecastpt <- unlist(lapply(varforecast$fcst, getfcst))
R> sum((varforecastpt - gdpVTSn2.0[52, ])^2)
\end{Sinput}
\begin{Soutput}
[1] 26.19805
\end{Soutput}
\end{Schunk}
\begin{table}[]
    \centering
    \begin{tabular}{lrr}\hline
        Model & \# Parameters & Prediction error \\ \hline
        GNAR$(2,[2,0])$ & 4 & 5.7 \\
        Individual AR$(2)$ & 38 & 8.1\\
        VAR$(1)$ & 199 & 26.2 \\ \hline
    \end{tabular}
    \caption{Estimated prediction error of differenced real GDP change at $t=52$ for all 35 countries.
	\label{tab:gdp}}
\end{table}
Our GNAR model gives a lower prediction error than both the AR and VAR results, reducing the error by 29\% compared to AR and by 78\% compared to VAR. Table~\ref{tab:gdp} summarises these results and also shows the number of
parameters fitted. It is clear that GNAR is particularly parsimonious.

We repeat the procedure above to perform analysis based upon two-step ahead forecasting. In this case, a different network minimises the prediction error for model GNAR(2,[2,2]). However, the BIC step identified that the GNAR(2,[0,0]) model had the best fit, which is a model that does not include network regression parameters.
\begin{Schunk}
\begin{Sinput}
R> gdpVTSn3 <- apply(gdpVTS, 2, function(x){x / sd(x[1:50], na.rm = TRUE)})
R> gdpPred <- predict(GNARfit(gdpVTSn2[1:50,], net = net921, alphaOrder = 2,
+    betaOrder = c(0, 0)), n.ahead=2)
R> sum((gdpPred[1,] - gdpVTSn3[51, ])^2)
\end{Sinput}
\begin{Soutput}
[1] 11.7874
\end{Soutput}
\begin{Sinput}
R> sum((gdpPred[2,] - gdpVTSn3[52, ])^2)
\end{Sinput}
\begin{Soutput}
[1] 8.067577
\end{Soutput}
\begin{Sinput}
R> arforecast <- apply(gdpVTSn3[1:50, ], 2, function(x){
+    forecast(auto.arima(x[!is.na(x)], d = 0, D = 0, max.p = 2, max.q = 0,
+      max.P = 0, max.Q = 0, stationary = TRUE, seasonal = FALSE, ic = "bic",
+      allowmean = FALSE, allowdrift = FALSE, trace = FALSE), h = 2)$mean
+  })
R> sum((arforecast[1,] - gdpVTSn3[51, ])^2)
\end{Sinput}
\begin{Soutput}
[1] 18.56074
\end{Soutput}
\begin{Sinput}
R> sum((arforecast[2,] - gdpVTSn3[52, ])^2)
\end{Sinput}
\begin{Soutput}
[1] 11.31722
\end{Soutput}
\begin{Sinput}
R> gdpVTSn3.0 <- gdpVTSn3
R> gdpVTSn3.0[is.na(gdpVTSn3.0)] <- 0
R> varforecast <- predict(restrict(VAR(gdpVTSn3.0[1:50, ], p = 1,
+    type = "none")), n.ahead = 2)
R> getfcst <- function(x){return(x[,1])}
R> varforecastpt <- matrix(unlist(lapply(varforecast$fcst, getfcst)),
+    nrow=2, ncol=35)
R> sum((varforecastpt[1,] - gdpVTSn3[51,])^2)
\end{Sinput}
\begin{Soutput}
[1] 114.9876
\end{Soutput}
\begin{Sinput}
R> sum((varforecastpt[2,] - gdpVTSn3[52,])^2)
\end{Sinput}
\begin{Soutput}
[1] 120.4467
\end{Soutput}
\end{Schunk}

Table~\ref{tab:gdp2steps} shows that the GNAR model is again the best performing, although in the two-step ahead prediction the fitted model is a special case of GNAR model with no neighbourhood parameters.
\begin{table}[]
    \centering
    \begin{tabular}{lrr}\hline
        Model & Prediction error at $t=51$ & Prediction error at $t=52$ \\ \hline
        GNAR$(2,[0,0])$ & 11.8 & 8.1 \\
        Individual AR$(2)$ & 18.6 & 11.3\\
        VAR$(1)$ & 115.0 & 120.4 \\ \hline
    \end{tabular}
    \caption{Estimated prediction error of differenced real GDP change at $t=51,52$, for all 35 countries.
	\label{tab:gdp2steps}}
\end{table}

Results in Tables~\ref{tab:gdp} and \ref{tab:gdp2steps} indicate that the VAR model works particularly poorly here, despite using thresholding to reduce the number of parameters. This example highlights that, for a multivariate series with many observations per time point, the VAR framework is restricted by the number of parameters that have to be fitted per time lag, thus reducing the
AR-order, $p$, it can capture. In addition, we were unable to find software to fit
 VAR models with for missing data  at the start of a series.

We end this section by noting that using Erd\H{o}s-R\'{e}nyi graphs are not the only type of network that could be used to aid prediction.  As suggested by a referee,
models such the Chung-Lu model \citep{aiello01:a, chung02:connected} could also be used to simulate random networks for this task; these graphs would allow for more
flexible network generation, for example using node-specific connection probabilities proportional to a country's size.

\section{Discussion and summary} \label{Discussion}
The \pkg{GNAR} package can be used to model network
time series using a network autoregressive structure.
Estimation under the proposed model is informed by the, potentially time-varying, structure of the network, assumed known.
Network time series models are in an early stage of development, but there is
enormous potential, especially as network data are increasingly being
collected and analysed in many fields. As far as possible, we attempt
to integrate our methods with existing valuable R functionality, such as its
linear modelling capability and the \code{fit} / \code{summary} / \code{predict} methods that are familiar
with \proglang{R} users.

Within our model a network is formed using edges of all covariates simultaneously, and the connection weights of this single network  can be calculated e.g., 
as described in Section~\ref{connectionweights}. Another approach is to consider a separate network for each covariate, and then calculate connection weights for each of
these networks. This would result in different  (known) weightings, $\omega$, and consequently  different fitted coefficients, $\beta$. The single network
approach is more appropriate for sparse networks and when different types of edge are closely related. In comparison, when covariates relate completely separate link
information between the nodes, use of different networks would be appropriate.

When covariates are present, the neighbour set structure is more complex, as different edge types can be included in a path between nodes. For example, in a network with
\emph{event} and \emph{proximal} edges,  network paths between stage-2 neighbours could include edges \emph{event-event}, \emph{event-proximal / proximal-event}, or
\emph{proximal-proximal}. These different types of path could be represented separately in the model using additional $\beta$ parameters. We note that the number of such
parameters would increase greatly for large covariate cardinality $C$ or high neighbour set stage $s_j$, so, in these cases, the large number of additional parameters may
not enhance the model.
Our model permits regression on any non-empty stage neighbour set, so models with high $s_j$ can be fitted.
For large $s_j$, the
neighbour sets may not be scientifically interpretable so small $s_j$ is recommended,
to favour parsimony and interpretability.

Trend is another factor that can seriously affect modelling and estimation, just as in
the regular time series situation. However, trend can be successfully modelled and estimated
by using second-generation wavelet (lifting) techniques before stochastic modelling,
as in \cite{Nunes2015}.

With the option of having different covariates and high order neighbourhood structures included, our GNAR model as presented in Section~\ref{Model} is incredibly flexible. In this article a sufficient condition for stationarity and consistency of the fitted parameters have been shown for the fixed network scenario. In addition, practical suggestions for order selection, and connection weights in the case of missing data have been discussed.


\bibliography{jss3594}

\begin{thebibliography}{49}
\newcommand{\enquote}[1]{``#1''}
\providecommand{\natexlab}[1]{#1}
\providecommand{\url}[1]{\texttt{#1}}
\providecommand{\urlprefix}{URL }
\expandafter\ifx\csname urlstyle\endcsname\relax
  \providecommand{\doi}[1]{doi:\discretionary{}{}{}#1}\else
  \providecommand{\doi}{doi:\discretionary{}{}{}\begingroup
  \urlstyle{rm}\Url}\fi
\providecommand{\eprint}[2][]{\url{#2}}

\bibitem[{Abegaz and Wit(2016)}]{abegaz16:sparseTSCGM}
Abegaz F, Wit E (2016).
\newblock \emph{\pkg{TSCGM}: Sparse Time Series Chain Graphical Models}.
\newblock \proglang{R}~package version~2.5,
  \urlprefix\url{http://CRAN.R-project.org/package=SparseTSCGM}.

\bibitem[{Adhikari \emph{et~al.}(2018)Adhikari, Junker, Sweet, and
  Thomas}]{adhikari18:HLSM}
Adhikari S, Junker B, Sweet T, Thomas AC (2018).
\newblock \emph{\pkg{HLSM}: Hierarchical Latent Space Network Model}.
\newblock \proglang{R}~package version~0.8,
  \urlprefix\url{http://CRAN.R-project.org/package=HLSM}.

\bibitem[{Aiello \emph{et~al.}(2001)Aiello, Chung, and Lu}]{aiello01:a}
Aiello W, Chung F, Lu L (2001).
\newblock \enquote{A Random Graph Model for Power Law Graphs.}
\newblock \emph{Experimental Mathematics}, \textbf{10}(1), 53--66.

\bibitem[{Akaike(1973)}]{akaike73:information}
Akaike H (1973).
\newblock \enquote{Information Theory and An Extension of the Maximum Likeihood
  Principle.}
\newblock In \emph{Proceedings of the 2nd International Symposium on
  Information Theory}, pp. 267--281. Academiai Kiado, Budapest, Hungary.

\bibitem[{Almutiry \emph{et~al.}(2018)Almutiry, Warriyar, and
  Deardon}]{almutiry18:EpiLMCT}
Almutiry V, Warriyar KV, Deardon R (2018).
\newblock \emph{\pkg{EpiLMCT}: Continuous Time Distance-Based and Network Based
  Individual Level Models for Epidemics}.
\newblock \proglang{R}~package version~1.1.2,
  \urlprefix\url{http://CRAN.R-project.org/package=EpiLMCT}.

\bibitem[{Bashir and Wei(2016)}]{Bashir16}
Bashir F, Wei HL (2016).
\newblock \enquote{Handling Missing Data in Multivariate Time Series Using a
  Vector Autoregressive Model Based Imputation (VAR-IM) Algorithm.}
\newblock In \emph{Proceedings of the 24th Mediterranean Conference on Control
  and Automation (MED)}, pp. 611--616. IEEE, Athens, Greece.

\bibitem[{Blocker \emph{et~al.}(2014)Blocker, Koullick, and
  Airoldi}]{blocker14:networkTomography}
Blocker AW, Koullick P, Airoldi E (2014).
\newblock \emph{\pkg{networkTomography}: Tools for Network Tomography}.
\newblock \proglang{R}~package version~0.3,
  \urlprefix\url{http://CRAN.R-project.org/package=networkTomography}.

\bibitem[{Brockwell and Davis(2006)}]{Brockwell2006}
Brockwell PJ, Davis RA (2006).
\newblock \emph{Time Series: Theory and Methods}.
\newblock 2nd edition. {Springer-Verlag}, New York.

\bibitem[{Brownlees(2017)}]{brownlees17:nets}
Brownlees C (2017).
\newblock \emph{\pkg{nets}: Network Estimation for Time Series}.
\newblock \proglang{R}~package version~0.9,
  \urlprefix\url{http://CRAN.R-project.org/package=nets}.

\bibitem[{Chung and Lu(2002)}]{chung02:connected}
Chung F, Lu L (2002).
\newblock \enquote{Connected Components in Random Graphs with Given Expected
  Degree Sequences.}
\newblock \emph{Annals of Combinatorics}, \textbf{6}(2), 125--145.

\bibitem[{Dahlhaus and Eichler(2003)}]{Dahlhaus2003}
Dahlhaus R, Eichler M (2003).
\newblock \enquote{Causality and Graphical Models for Time Series.}
\newblock In PJ~Green, NL~Hjort, S~Richardson (eds.), \emph{Highly Structured
  Stochastic Systems}, pp. 115--137. Oxford University Press, Oxford.

\bibitem[{Denny \emph{et~al.}(2018)Denny, Wilson, Cranmer, Desmarais, and
  Bhamidi}]{denny18:GERGM}
Denny MJ, Wilson JD, Cranmer S, Desmarais BA, Bhamidi S (2018).
\newblock \emph{\pkg{GERGM}: Estimation and Fit Diagnostics for Generalized
  Exponential Random Graph Models}.
\newblock \proglang{R}~package version~0.13.0,
  \urlprefix\url{http://CRAN.R-project.org/package=GERGM}.

\bibitem[{Emerencia(2018)}]{emerencia18:autovarCore}
Emerencia A (2018).
\newblock \emph{\pkg{autovarCore}: Automated Vector Autoregression Models and
  Networks}.
\newblock \proglang{R}~package version~1.0-4,
  \urlprefix\url{http://CRAN.R-project.org/package=autovarCore}.

\bibitem[{Epskamp(2018)}]{epskamp18:graphicalVAR}
Epskamp S (2018).
\newblock \emph{\pkg{graphicalVAR}: Graphical VAR for Experience Sampling
  Data}.
\newblock \proglang{R}~package version~0.2.2,
  \urlprefix\url{http://CRAN.R-project.org/package=graphicalVAR}.

\bibitem[{Epskamp \emph{et~al.}(2019)Epskamp, Deserno, and
  Bringmann}]{epskamp19:mlVAR}
Epskamp S, Deserno MK, Bringmann LF (2019).
\newblock \emph{\pkg{mlVAR}: Multi-Level Vector Autoregression}.
\newblock \proglang{R}~package version~0.4.2,
  \urlprefix\url{http://CRAN.R-project.org/package=mlVAR}.

\bibitem[{Grimmett(2010)}]{Grimmett2010}
Grimmett G (2010).
\newblock \emph{Probability on Graphs: Random Processes on Graphs and
  Lattices}.
\newblock Cambridge University Press, Cambridge.

\bibitem[{Groendyke \emph{et~al.}(2018)Groendyke, Welch, and
  Hunter}]{groendyke18:epinet}
Groendyke C, Welch D, Hunter D (2018).
\newblock \emph{\pkg{epinet}: Epidemic/Network-Related Tools}.
\newblock \proglang{R}~package version~2.1.8,
  \urlprefix\url{http://CRAN.R-project.org/package=epinet}.

\bibitem[{Guerrero and Gaspar(2010)}]{Guerrero10}
Guerrero V, Gaspar B (2010).
\newblock \enquote{Edition and Imputation of Multiple Time Series Data
  Generated by Repetitive Surveys.}
\newblock \emph{Journal of Data Science}, \textbf{8}, 555--577.

\bibitem[{Handcock \emph{et~al.}(2018)Handcock, Hunter, Butts
  \emph{et~al.}}]{handcock18:ergm}
Handcock MS, Hunter DR, Butts CT, \emph{et~al.} (2018).
\newblock \emph{\pkg{ergm}: Fit, Simulate and Diagnose Exponential-Family
  Models for Networks}.
\newblock \proglang{R}~package version~3.9.4,
  \urlprefix\url{http://CRAN.R-project.org/package=ergm}.

\bibitem[{Haslbeck(2019)}]{haslbeck19:mgm}
Haslbeck J (2019).
\newblock \emph{\pkg{mgm}: Estimating Time-Varying k-Order Mixed Graphical
  Models}.
\newblock \proglang{R}~package version~1.2-6,
  \urlprefix\url{http://CRAN.R-project.org/package=mgm}.

\bibitem[{Honaker and King(2010)}]{Honaker10}
Honaker J, King G (2010).
\newblock \enquote{What to Do about Missing Values in Time-Series Cross-Section
  Data.}
\newblock \emph{American Journal of Political Science}, \textbf{54}, 561--581.

\bibitem[{Hyndman \emph{et~al.}(2017)Hyndman, Athanasopoulos,
  Christoph~Bergmeir, Caceres \emph{et~al.}}]{hyndman17:forecast}
Hyndman R, Athanasopoulos G, Christoph~Bergmeir C, Caceres G, \emph{et~al.}
  (2017).
\newblock \emph{\pkg{forecast}: Forecasting Functions for Time Series and
  Linear Models}.
\newblock \proglang{R}~package version~8.0,
  \urlprefix\url{http://CRAN.R-project.org/package=forecast}.

\bibitem[{Hyndman and Khandakar(2008)}]{Hyndman2008}
Hyndman RJ, Khandakar Y (2008).
\newblock \enquote{Automatic Time Series Forecasting: The \pkg{forecast}
  package for \proglang{R}.}
\newblock \emph{Journal of Statistical Software}, \textbf{27}(3).

\bibitem[{{Knight} \emph{et~al.}(2016){Knight}, {Nunes}, and
  {Nason}}]{Knight2016}
{Knight} MI, {Nunes} MA, {Nason} GP (2016).
\newblock \enquote{{Modelling, Detrending and Decorrelation of Network Time
  Series}.}
\newblock \emph{ArXiv e-prints}, \textbf{1603.03221}.

\bibitem[{Kolaczyk(2009)}]{Kolaczyk2009}
Kolaczyk E (2009).
\newblock \emph{Statistical Analysis of Network Data: Methods and Models}.
\newblock Springer-Verlag, New York.

\bibitem[{Krvitsky \emph{et~al.}(2018{\natexlab{a}})Krvitsky, Handcock, Hunter,
  Goodreau \emph{et~al.}}]{krvitsky18:tergm}
Krvitsky P, Handcock MS, Hunter DR, Goodreau SM, \emph{et~al.}
  (2018{\natexlab{a}}).
\newblock \emph{\pkg{tergm}: Fit, Simulate and Diagnose Models for Network
  Evolution Based on Exponential-Family Random Graph Models}.
\newblock \proglang{R}~package version~3.5.2,
  \urlprefix\url{http://CRAN.R-project.org/package=tergm}.

\bibitem[{Krvitsky \emph{et~al.}(2018{\natexlab{b}})Krvitsky, Handcock,
  Shortreed, Tantrum \emph{et~al.}}]{krvitsky18:latentnet}
Krvitsky P, Handcock MS, Shortreed SM, Tantrum J, \emph{et~al.}
  (2018{\natexlab{b}}).
\newblock \emph{\pkg{latentnet}: Latent Position and Cluster Models for
  Statistical Networks}.
\newblock \proglang{R}~package version~2.9.0,
  \urlprefix\url{http://CRAN.R-project.org/package=latentnet}.

\bibitem[{Lane \emph{et~al.}(2019)Lane, Gates, Fisher, Molenaar
  \emph{et~al.}}]{lane19:gimme}
Lane S, Gates K, Fisher Z, Molenaar P, \emph{et~al.} (2019).
\newblock \emph{\pkg{gimme}: Group Iterative Multiple Model Estimation}.
\newblock \proglang{R}~package version~0.5-1,
  \urlprefix\url{http://CRAN.R-project.org/package=gimme}.

\bibitem[{Leeming \emph{et~al.}(2019)Leeming, Nason, Nunes, and
  Knight}]{leeming18:GNAR}
Leeming K, Nason GP, Nunes MA, Knight MI (2019).
\newblock \emph{\pkg{GNAR}: Methods for Fitting Network Time Series Models}.
\newblock \proglang{R}~package version~1.0,
  \urlprefix\url{http://CRAN.R-project.org/package=GNAR}.

\bibitem[{Leeming(2019)}]{LeemingPhD}
Leeming KA (2019).
\newblock \emph{New Methods in Time Series Analysis: Univariate Testing and
  Network Autoregression Modelling}.
\newblock Ph.D. thesis, University of Bristol.

\bibitem[{Leger(2015)}]{leger15:blockmodels}
Leger JB (2015).
\newblock \emph{\pkg{blockmodels}: Latent and Stochastic Block Model Estimation
  by a `V-EM' Algorithm}.
\newblock \proglang{R}~package version~1.1.1,
  \urlprefix\url{http://CRAN.R-project.org/package=blockmodels}.

\bibitem[{Leifeld and Cranmer(2017)}]{leifeld17:tnam}
Leifeld P, Cranmer SJ (2017).
\newblock \emph{\pkg{tnam}: Temporal Network Autocorrelation Models (TNAM)}.
\newblock \proglang{R}~package version~1.6.5,
  \urlprefix\url{http://CRAN.R-project.org/package=tnam}.

\bibitem[{L\"{u}tkepohl(2005)}]{Lutkepohl2005}
L\"{u}tkepohl H (2005).
\newblock \emph{New Introduction to Multiple Time Series Analysis}.
\newblock Springer-Verlag, Berlin.

\bibitem[{Manitz and Harbering(2018)}]{manitz18:NetOrigin}
Manitz J, Harbering J (2018).
\newblock \emph{\pkg{NetOrigin}: Origin Estimation for Propagation Processes on
  Complex Networks}.
\newblock \proglang{R}~package version~1.0-3,
  \urlprefix\url{http://CRAN.R-project.org/package=NetOrigin}.

\bibitem[{Manyika \emph{et~al.}(2016)Manyika, Lund, Bughin, Woetzel, Stamenov,
  and Dhingra}]{McKinsey2016}
Manyika J, Lund S, Bughin J, Woetzel J, Stamenov K, Dhingra D (2016).
\newblock \enquote{Digital Globalization: The New Era of Global Flows.}
\newblock McKinsey, London.

\bibitem[{Marquez \emph{et~al.}(2018)Marquez, Grisi-Filho, and
  Amaku}]{marquez18:hybridModels}
Marquez FS, Grisi-Filho JHH, Amaku M (2018).
\newblock \emph{\pkg{hybridModels}: Stochastic Hybrid Models in Dynamic
  Networks}.
\newblock \proglang{R}~package version~0.3.5,
  \urlprefix\url{http://CRAN.R-project.org/package=hybridModels}.

\bibitem[{Matias and Miele(2018)}]{matias18:dynsbm}
Matias C, Miele V (2018).
\newblock \emph{\pkg{dynsbm}: Dynamic Stochastic Block Models}.
\newblock \proglang{R}~package version~0.5,
  \urlprefix\url{http://CRAN.R-project.org/package=dynsbm}.

\bibitem[{Nunes \emph{et~al.}(2015)Nunes, Knight, and Nason}]{Nunes2015}
Nunes MA, Knight MI, Nason GP (2015).
\newblock \enquote{Modelling And Prediction of Time Series Arising On a Graph.}
\newblock In A~Antoniadis, JM~Poggi, X~Brossat (eds.), \emph{Modeling and
  Stochastic Learning for Forecasting in High Dimensions}, volume 217 of
  \emph{Lecture Notes in Statistics}, pp. 183--192. Springer-Verlag, New York.

\bibitem[{Pfaff(2008)}]{Pfaff2008}
Pfaff B (2008).
\newblock \enquote{VAR, SVAR and SVEC Models: Implementation Within
  \proglang{R} Package \pkg{vars}.}
\newblock \emph{Journal of Statistical Software}, \textbf{27}(4).

\bibitem[{{\proglang{R} Core Team}(2019)}]{Rcore2017}
{\proglang{R} Core Team} (2019).
\newblock \emph{\proglang{R}: A Language and Environment for Statistical
  Computing}.
\newblock \proglang{R} Foundation for Statistical Computing, Vienna, Austria.
\newblock \urlprefix\url{https://www.R-project.org/}.

\bibitem[{Schwarz(1978)}]{Schwarz1978}
Schwarz G (1978).
\newblock \enquote{Estimating the Dimension of a Model.}
\newblock \emph{The Annals of Statistics}, \textbf{6}, 461--464.

\bibitem[{Schweinberger \emph{et~al.}(2018)Schweinberger, Handcock, and
  Luna}]{schweinberger18:hergm}
Schweinberger M, Handcock MS, Luna P (2018).
\newblock \emph{\pkg{hergm}: Hierarchical Exponential-Family Random Graph
  Models}.
\newblock \proglang{R}~package version~3.2-1,
  \urlprefix\url{http://CRAN.R-project.org/package=hergm}.

\bibitem[{Spencer \emph{et~al.}(2015)Spencer, Hill, and
  Mukherjee}]{Spencer2015}
Spencer SEF, Hill SM, Mukherjee S (2015).
\newblock \enquote{Inferring Network Structure from Interventional Time-Course
  Experiments.}
\newblock \emph{The Annals of Statistics}, \textbf{9}, 507--524.

\bibitem[{Tsay(2014)}]{Tsay2014}
Tsay RS (2014).
\newblock \emph{Multivariate Time Series Analysis}.
\newblock John Wiley \& Sons, Hoboken.

\bibitem[{Varga(1962)}]{Varga1962}
Varga RS (1962).
\newblock \emph{Matrix Iterative Analysis}.
\newblock Prentice-Hall, New Jersey.

\bibitem[{Vazoller \emph{et~al.}(2016)Vazoller, Frattarolo, and
  Billio}]{vazoller16:sparsevar}
Vazoller S, Frattarolo L, Billio M (2016).
\newblock \emph{\pkg{sparsevar}: A Package for Sparse VAR/VECM Estimation}.
\newblock \proglang{R}~package version~0.0.10,
  \urlprefix\url{http://CRAN.R-project.org/package=sparsevar}.

\bibitem[{Warriyar and Deardon(2018)}]{warriyar18:EpiLM}
Warriyar KV, Deardon R (2018).
\newblock \emph{\pkg{EpiLM}: Spatial and Network Based Individual Level Models
  for Epidemics}.
\newblock \proglang{R}~package version~1.4.2,
  \urlprefix\url{http://CRAN.R-project.org/package=EpiLM}.

\bibitem[{Wyse \emph{et~al.}(2017)Wyse, Ryan, and Friel}]{wyse17:collpcm}
Wyse J, Ryan C, Friel N (2017).
\newblock \emph{\pkg{collpcm}: Collapsed Latent Position Cluster Model for
  Social Networks}.
\newblock \proglang{R}~package version~1.0,
  \urlprefix\url{http://CRAN.R-project.org/package=collpcm}.

\bibitem[{Zhu \emph{et~al.}(2017)Zhu, Pan, Li, Liu, and Wang}]{Zhu2017}
Zhu X, Pan R, Li G, Liu Y, Wang H (2017).
\newblock \enquote{Network Vector Autoregression.}
\newblock \emph{The Annals of Statistics}, \textbf{45}, 1096--1123.

\end{thebibliography}
\appendix
\section{Proof of stationarity conditions for the GNAR model} \label{appstatcon}
A sufficient condition for stationarity of the GNAR model (\ref{NAR1}) with a static network is
\begin{equation}
\sum_{j=1}^p \left( |\alpha_{i,j}| +  \sum\limits_{c=1}^C\sum\limits_{r=1}^{s_j}|\beta_{j,r,c}| \right)<1     \quad \forall i \in 1,...,N.
\end{equation} \bigskip \\
\textit{Proof:}
First Gerschgorin's theorem and a corollary are presented without proof, both taken from \cite{Varga1962}.

\noindent\rule{\textwidth}{1pt}

\textbf{Theorem} Let $\mathit{A}=(a_{i,j})$ be an arbitrary $n\times n$ complex matrix, and let $\Lambda_i \equiv \sum\limits_{j=1, j\neq i}^n |a_{i,j}|$, $1\leq i \leq n$. Then, all of the eigenvalues $\lambda$ of $\mathit{A}$ lie in the union of the disks $|z-a_{i,i}| \leq \Lambda_i$, $1\leq i \leq n$.

Since the disk $|z-a_{i,i}| \leq \Lambda_i$ is a subset of the disk $|z| \leq |a_{i,i}| + \Lambda_i$, we have the immediate result of \\
\textbf{Corollary 1} If $\mathit{A}=(a_{i,j})$ is an arbitrary $n \times n$ complex matrix with eigenvalues $\lambda_i$, $1\leq i \leq n$, and $\nu \equiv \max\limits_{1 \leq i \leq n} \sum\limits_{j=1}^n |a_{i,j}|$, then $\max\limits_{1\leq i \leq n} |\lambda_i| \leq \nu$.\\

\noindent\rule{\textwidth}{1pt}

We can write the static-network GNAR process
$\mathbf{X}_t=(X_{1,t},...,X_{N,t})'$ as a VAR process, by  writing
$\mathbf{X}_t=\mathit{\phi}_1 \mathbf{X}_{t-1} + ... +\mathit{\phi}_p \mathbf{X}_{t-p} + \mathbf{u}_t$, where $\mathit{\phi}_k$ are $n\times n$ matrices such that $\mathit{\phi}_k=\diag\{\alpha_{i,k}\} + \sum\limits_{c=1}^C \sum_{r=1}^{s_k} \beta_{k,r,c} \mathit{W}^{(r,c)}$,  where matrices $\mathit{W}^{(r,c)}$ have entries $[\mathit{W}^{(r,c)}]_{\ell,m}=\omega_{\ell,m,c} \mathbb{I}\{m \in \mathcal{N}^{(r)}(\ell)\}$ and $\mathbf{u}_t$ is the vector of errors at time $t$. We use the notation $[\cdot]_{\ell,m}$ to denote the $\ell,m$ entry of a matrix.

From~\cite{Brockwell2006}, for example,
we have that if $\det (I_N-\mathit{\phi}_1z-...-\mathit{\phi}_pz^p) \neq 0$, for all $z \in \mathbb{C}$ such that $|z|\leq 1$, then the VAR model has exactly one stationary solution. Using Lemma~2.1 from \cite{Tsay2014} we have $\det (I_N-\mathit{\phi}_1z-...-\mathit{\phi}_pz^p) = \det(I_{Np} - \varPhi z) $, where $\varPhi$ is the $Np\times Np$ companion matrix defined as
\begin{equation*}
\varPhi=
\begin{bmatrix}
    \mathit{\phi}_1 & \mathit{\phi}_2 & \hdots & \mathit{\phi}_{p-1} & \mathit{\phi}_p \\
    \mathit{I}_N & \mathit{0}_N & \hdots & \mathit{0}_N & \mathit{0}_N \\
    \mathit{0}_N & \mathit{I}_N & \hdots & \mathit{0}_N & \mathit{0}_N \\
    \vdots & \vdots & \ddots & \vdots & \vdots \\
    \mathit{0}_N & \mathit{0}_N &  \hdots & \mathit{I}_N & \mathit{0}_N \\
\end{bmatrix},
\end{equation*}
where $\mathit{I}_N$ and $\mathit{0}_N$ are the $N\times N$ identity and zero matrices, respectively.\footnote{Note that $\varPhi$ is defined differently in the two books, this is the Tsay (2014) version.} Thus we require that the roots of $\det(\mathit{I}_{Np} - \varPhi z)$ are outside of the unit circle for stationarity, or equivalently, that the eigenvalues of $\varPhi$ lie inside the unit circle.

\noindent
We investigate the eigenvalues of $\varPhi$ using Corollary 1.

\noindent
For rows $N+1, \ldots, Np$, $\max\limits_{N+1 \leq \ell \leq Np} \sum\limits_{m=1}^{Np} |\varPhi_{\ell, m}| = 1$. \\
For rows $1, \ldots, N$,
\begin{equation*}
\begin{split}
    \max\limits_{1 \leq \ell \leq N} \sum\limits_{m=1}^{Np} |\varPhi_{\ell, m}| &=
        \max\limits_{1 \leq \ell \leq N} \sum\limits_{s=1}^{N} \sum\limits_{k=1}^p \abs*{\left[ \phi_k \right]_{\ell, s}} \\
    &= \max\limits_{1 \leq \ell \leq N} \sum\limits_{s=1}^{N} \sum\limits_{k=1}^p \abs*{\left[ \diag \{\alpha_{i,k}\}+\sum\limits_{c=1}^C\sum\limits_{r=1}^{s_k} \beta_{k,r,c}\mathit{W}^{(r,c)} \right]_{\ell, s}} \\
    &\leq \max\limits_{1 \leq \ell \leq N} \sum\limits_{s=1}^{N} \sum\limits_{k=1}^p \left[ \diag \{\abs*{\alpha_{i,k}}\}+\sum\limits_{c=1}^C\sum\limits_{r=1}^{s_k} \abs*{ \beta_{k,r,c}}\mathit{W}^{(r,c)} \right]_{\ell, s} \\
    &= \max\limits_{1 \leq \ell \leq N} \sum\limits_{s=1}^{N} \sum\limits_{k=1}^p  \Bigg(\abs*{\alpha_{\ell,k}}\mathbb{I}\{\ell=s\} \\
    & \quad \quad \quad \quad \quad  + \sum\limits_{c=1}^C\sum\limits_{r=1}^{s_k} \abs*{ \beta_{k,r,c}}\omega_{\ell,s,c}\mathbb{I}\{s \in \mathcal{N}^{(r)}(\ell)\} \Bigg) \\
    &= \max\limits_{1 \leq \ell \leq N}  \sum\limits_{k=1}^p  \Bigg(\abs*{\alpha_{\ell,k}}\sum\limits_{s=1}^{N}\mathbb{I}\{\ell=s\} \\
    & \quad \quad \quad \quad \quad  +\sum\limits_{c=1}^C\sum\limits_{r=1}^{s_k} \abs*{ \beta_{k,r,c}}\sum\limits_{s=1}^{N}\omega_{\ell,s,c}\mathbb{I}\{s \in \mathcal{N}^{(r)}(\ell)\} \Bigg) \\
    &\leq \max\limits_{1 \leq \ell \leq N} \sum\limits_{k=1}^p  \left(\abs*{\alpha_{\ell,k}}+\sum\limits_{c=1}^C\sum\limits_{r=1}^{s_k} \abs*{ \beta_{k,r,c}} \right),
\end{split}
\end{equation*}
as at each node $\ell \in \mathcal{K}$ and each covariate $c \in \{1,\hdots,C\}$, $\sum\limits_{s \in \mathcal{N}^{(r)}(\ell)} \omega_{\ell,s,c} \leq 1$. Under condition (\ref{statcon}), $\max\limits_{1 \leq \ell \leq N} \sum\limits_{m=1}^{Np} |\varPhi_{\ell, m}| < 1$. Therefore, $\max\limits_{1 \leq \ell \leq Np} \sum\limits_{m=1}^{Np} |\varPhi_{\ell, m}| \leq 1$, and, using Corollary 1, we have that the spectral radius of $\varPhi$ is at most one.

\noindent
We next check whether an eigenvalue with modulus 1 is possible.

Assume that there exists an eigenvalue, $\lambda$ of $\mathit{\Phi}$ such that $|\lambda| =1$. By definition, there exists an eigenvector $\mathbf{v} \in \mathbb{C}^{Np}$ such that $\mathit{\Phi}\mathbf{v}=\lambda\mathbf{v}$. By writing $\mathbf{v}=(\mathbf{v}_1',\hdots, \mathbf{v}_p')'$, where each $\mathbf{v}_k$ is a column vector of length $N$, we can rewrite the eigenequation as the following simultaneous equations:
\begin{equation}
\text{(i)} \ \sum\limits_{k=1}^p \mathit{\phi}_k \mathbf{v}_k =\lambda \mathbf{v}_1 \text{ and (ii) }
\mathbf{v}_k= \lambda \mathbf{v}_{k+1}, \forall k \in \{1,\hdots,p-1\}.
\end{equation}
Therefore $\mathbf{v}_k=\lambda^{p-k}\mathbf{v}_p \quad \forall k \in \{1,\hdots,p\}$ and, replacing this on both sides of (i), we have that $\sum_{k=1}^p \mathit{\phi}_k \lambda^{p-k} \mathbf{v}_p = \lambda^p \mathbf{v}_p$. This results in the equation $\sum_{k=1}^p \mathit{\phi}_k \lambda^{-k} \mathbf{v}_p = \mathbf{v}_p$, which can be written in matrix form as $\varPsi \mathbf{v}_p = \mathbf{v}_p$, where $\varPsi$ is the $N\times N$ matrix $\varPsi =\sum_{k=1}^p \mathit{\phi}_k \lambda^{-k} $.

\noindent
Hence, if $\varPhi$ has an eigenvalue of modulus 1, then $\varPsi$ must have 1 as an eigenvalue.

\noindent
We again use Corollary 1 for the eigenvalues of $\varPsi$, under the assumption $\abs*{\lambda}=1$.
\begin{equation*}
\begin{split}
    \max\limits_{1 \leq \ell \leq N} \sum\limits_{m=1}^N \abs*{\varPsi_{\ell, m} } &= \max\limits_{1 \leq \ell \leq N} \sum\limits_{m=1}^N \abs*{\left[\sum\limits_{k=1}^p \mathit{\phi}_k \lambda^{k-2} \right]_{\ell, m} } \\
    &= \max\limits_{1 \leq \ell \leq N} \sum\limits_{m=1}^N \abs*{\sum\limits_{k=1}^p \lambda^{k-2} \left[ \diag\{\alpha_{i,k}\}  +\sum\limits_{c=1}^C\sum\limits_{r=1}^{s_k} \beta_{k,r,c}  \mathit{W}^{(r,c)}\right]_{\ell, m} } \\
    &\leq \max\limits_{1 \leq \ell \leq N} \sum\limits_{m=1}^N \sum\limits_{k=1}^p \abs*{\lambda^{k-2}} \Bigg( \abs*{\alpha_{\ell,k}}\mathbb{I}\{\ell = m\} \\
    & \quad \quad \quad \quad +\sum\limits_{c=1}^C\sum\limits_{r=1}^{s_k} \abs*{\beta_{k,r,c}}  \omega_{\ell,m,c} \mathbb{I}\{m \in \mathcal{N}^{(r)}(\ell) \}\Bigg) \\
    &= \max\limits_{1 \leq \ell \leq N}  \sum\limits_{k=1}^p \left( \abs*{\alpha_{\ell,k}} +\sum\limits_{c=1}^C\sum\limits_{r=1}^{s_k} \abs*{\beta_{k,r,c}} \sum\limits_{m=1}^N \omega_{\ell,m,s}\mathbb{I}\{m \in \mathcal{N}^{(r)}(\ell) \} \right) \\
    &\leq \max\limits_{1 \leq \ell \leq N} \sum\limits_{c=1}^C\sum\limits_{k=1}^p \left( \abs*{\alpha_{\ell,k}} +\sum\limits_{c=1}^C\sum\limits_{r=1}^{s_k} \abs*{\beta_{k,r,c}}  \right) \\
    \end{split}
\end{equation*}
Under condition (\ref{statcon}) this is smaller than 1, so, by Corollary 1, no eigenvalues of $\varPsi$ have modulus 1 or greater. This contradicts the assumption that an eigenvalue of $\varPhi$, $\lambda$, exists such that $|\lambda |=1$. Hence, the eigenvalues of $\varPhi$ are inside the unit circle under condition (\ref{statcon}) and the GNAR model is stationary.

\section{Parameter estimate consistency}\label{consistency}
We employ least squares estimation for the GNAR model parameters and establish their consistency using results from \cite{Lutkepohl2005}. The column form of the static-network GNAR$(p,[\mathbf{s}])$ model can be written in a VAR framework as
\begin{equation*}
    \mathbf{X}_t = \mathit{\phi}_1 \mathbf{X}_{t-1} + \hdots + \mathit{\phi}_p \mathbf{X}_{t-p} + \mathbf{u}_t,
\end{equation*} where the matrices $\mathit{\phi}_i$ contain the network information. In matrix form the GNAR model is $\mathit{X}=\mathit{BZ} + \mathit{U}$, where $\mathit{X}=[\mathbf{X}_{p+1},\hdots,\mathbf{X}_T]$, $\mathit{B}=[\mathit{\phi}_1,\hdots,\mathit{\phi}_p]$, $\mathit{Z}=[\mathbf{Z}_p,\hdots,\mathbf{Z}_{T-1}]$, with $\mathbf{Z}'_t= [
    \mathbf{X}_t,
    \hdots,
    \mathbf{X}_{t-p+1}]$, and $\mathit{U} = [\mathbf{u}_{p+1},\hdots,\mathbf{u}_T]$.
The constraints imposed to form a GNAR model can be written linearly as $\vect{(\mathit{B})} = \mathit{R}\boldsymbol{\gamma}$, where $\mathit{R}$ is the constraint matrix embedding the network structure of dimension $pN^2 \times M$, $\boldsymbol{\gamma}$ is an unrestricted parameter vector of length $M$, where $M$ is defined as in Section~\ref{BIC}, and $\vect$ is the operator that stacks the columns of a matrix into a vector.
Using the estimated generalised least squares estimator, we apply  results from Section~5.2 of \cite{Lutkepohl2005} to obtain consistency for the GNAR parameters.
Let $\otimes$ denote the Kronecker product and $\plim$ denote limit in probability.
\begin{proposition}
Suppose $\{\mathbf{X}_t\}$ is an $N$-dimensional, stationary GNAR$(p)$ process with a static network, whose innovations $\{\mathbf{u}_t\}$ are independent white noise with finite fourth moment, and covariance matrix ${\varSigma}_u$. \\
Then, given an estimator of the innovation covariance matrix $\tilde{{\varSigma}}_u$, such that $\plim \tilde{{\varSigma}}_u = {\varSigma}_u$, the estimated generalised least squares estimator of the unrestricted parameters,
\begin{equation*}{\tilde{\boldsymbol{\gamma}} = \{\mathit{R'(Z Z}' \otimes \tilde{\varSigma}_u^{-1}) \mathit{R}\}^{-1} \mathit{R} (\mathit{Z} \otimes \tilde{\varSigma}_u^{-1} ) \vect (\mathit{X})},
\end{equation*}
 is consistent; ${\plim \tilde{\boldsymbol{\gamma}} = \boldsymbol{\gamma}}$ and ${\sqrt{T} (\tilde{\boldsymbol{\gamma}} -\boldsymbol{\gamma}) \rightarrow^d N[0, \{\mathit{R}' (\varGamma \otimes \tilde{{\varSigma}}_u^{-1} )R\}^{-1} ]}$ where ${\varGamma=\plim T^{-1} \mathit{Z Z}' }$.
\end{proposition}
\noindent Again, adapting \cite{Lutkepohl2005}, we have the following result for a consistent estimator of the innovation covariance matrix in the GNAR setting.
\begin{proposition}
A consistent estimator of $\varSigma_u$ is given by
\begin{displaymath}
{\tilde{\varSigma}_u = T^{-1} (\mathit{X}-\hat{\mathit{B}}\mathit{Z})(\mathit{X}-\hat{\mathit{B}}\mathit{Z})'},
\end{displaymath}
where $\hat{\mathit{B}}\mathit{Z}$ are the fitted values from estimating the parameters using the least squares estimator ${\hat{\boldsymbol{\gamma}} = \{\mathit{R}'(\mathit{ZZ}' \otimes \mathit{I}_N)\mathit{R}\}^{-1} \mathit{R}' (\mathit{Z} \otimes \mathit{I}_N) \vect(\mathit{X})}$.
\end{proposition}
\noindent Estimating the parameters with $\hat{\boldsymbol{\gamma}}$ involves using the linear constraints, but assumes independent and identically distributed innovations across nodes.

\section{Further GNAR model fitting examples}\label{extrasims}
For the data in Section~\ref{exsim}, we could fit a model using individual alpha parameters, i.e., a GNAR(1, [1]):

\begin{Schunk}
\begin{Sinput}
R> print(GNARfit(vts = fiveVTS, net = fiveNet, alphaOrder = 1,
+    betaOrder = 1, globalalpha = FALSE))
\end{Sinput}
\begin{Soutput}
Model: 
GNAR(1,[1]) 

Call:
lm(formula = yvec ~ dmat + 0)

Coefficients:
dmatalpha1node1  dmatalpha1node2  dmatalpha1node3  dmatalpha1node4  
        0.03884          0.23248          0.21101          0.18413  
dmatalpha1node5      dmatbeta1.1  
        0.23273          0.48764  
\end{Soutput}
\end{Schunk}
An alternative model could separate the nodes A and B to have different parameters than C, D, and E:
\begin{Schunk}
\begin{Sinput}
R> print(GNARfit(vts = fiveVTS, net = fiveNet, alphaOrder = 1,
+    betaOrder = 1, fact.var = c("AB", "AB", "CDE", "CDE", "CDE")))
\end{Sinput}
\begin{Soutput}
Model: 
GNAR(1,[1]) 

Call:
lm(formula = yvec ~ dmat + 0)

Coefficients:
  dmatalpha1 'AB'   dmatbeta1.1 'AB'   dmatalpha1 'CDE'  
           0.1749             0.3901             0.1903  
dmatbeta1.1 'CDE'  
           0.5652  
\end{Soutput}
\end{Schunk}

\end{document}